\documentclass[a4paper,fleqn,usenatbib]{mnras}

\usepackage[T1]{fontenc}
\usepackage{ae,aecompl}

\usepackage{graphicx}	% Including figure files
\usepackage{graphics}
\usepackage{amsmath}	% Advanced maths commands
\usepackage{amssymb}	% Extra maths symbols

\title[A supersoft X-ray source in M31]{CXO J004318.8+412016, a steady supersoft X-ray source in M 31}

\author[M. Orio et al.]{
Marina Orio,$^{1,2}$\thanks{E-mail: orio@astro.wisc.edu}
G. J. M. Luna,$^{3,4,5}$
R. Kotulla,$^{1}$
J. S. Gallager,$^{1}$
L. Zampieri,$^{2}$
\newauthor
J. Mikolajewska,$^{6}$
D. Harbeck,$^{7}$
A.Bianchini,$^{8}$
E. Chiosi,$^{8}$
M. Della Valle,$^{9,10}$
\newauthor
D. de Martino,$^{9}$
A. Kaur,$^{11}$
M. Mapelli,$^{2}$
U. Munari,$^{2}$
A. Odendaal,$^{12}$ 
G. Trinchieri,$^{13}$
\newauthor
J. Wade,$^1$
P. Zemko$^{8}$\\
$^1$ Department of Astronomy, University of Wisconsin, 475 N. Charter Str., Madison WI 53704\\
$^2$ INAF--Osservatorio di Padova, vicolo dell' Osservatorio 5,
   I-35122 Padova, Italy \\
$^{3}$Universidad de Buenos Aires, Facultad de Ciencias Exactas y Naturales, Buenos Aires, Argentina\\ 
$^{4}$CONICET-Universidad de Buenos Aires, Instituto de Astronom\'ia y F\'isica del Espacio, (IAFE), Av. Inte. G\"uiraldes 2620, C1428ZAA, Buenos Aires, Argentina\\
$^{5}$ Universidad Nacional Arturo Jauretche, Av. Calchaqu\'i 6200, F. Varela, Buenos Aires, Argentina \\
$^6$ N. Copernicus Astronomical Center, Polish Academy of Sciences, Bartycka 18, PL 00-716 Warsaw, Poland
\\
$^7$ WIYN Observatory, Tucson, AZ 85719, USA\\
$^8$ Dipartimento di Astronomia, Universit\`a di Padova, vicolo
 Osservatorio, 2, 35122 Padova, Italy \\ 
$^9$ INAF-Osservatorio Astronomico di Capodimonte, Salita Moiariello 16, I-80131 Napoli \\
$^{10}$ International Center for Relativistic Astrophysics, Piazzale della Repubblica, 2, 65122, Pescara, Italy\\
$^{11}$ Department of Physics and Astronomy, Clemson University, Clemson, SC, 29634, USA \\
$^{12}$ Department of Physics, University of the Free State, PO Box 339, Bloemfontein 9300, South Africa \\
$^{13}$ INAF-Osservatorio Astronomico di Brera, via Brera 28, 20121 Milano 
}
\date{Accepted XXX. Received YYY; in original form ZZZ}

\pubyear{2016}

\begin{document}
\label{firstpage}
\pagerange{\pageref{firstpage}--\pageref{lastpage}}
\maketitle

\begin{abstract}
We obtained an optical spectrum of a star we identify as  the optical 
 counterpart of the M31 Chandra source CXO J004318.8+412016,
because of  prominent emission lines of the Balmer series,
of neutral helium, and  a He II line at 4686 \AA.
The continuum energy distribution and the spectral characteristics
 demonstrate the presence of a red giant of
 K or earlier spectral type, so we concluded that the binary is
 likely to be a symbiotic system.  CXO J004318.8+412016 has been
 observed in X-rays as a luminous supersoft source (SSS) since 1979, 
 with effective temperature exceeding 40 eV and
 variable X-ray luminosity, oscillating between
 a few times 10$^{35}$ erg  s$^{-1}$
 and a few times 10$^{37}$ erg s$^{-1}$ in the space of a few 
 weeks.  The optical, infrared and ultraviolet
 colors of the optical object are consistent with 
 an an accretion disk around
 a compact object companion, which may either be a white dwarf, or a black hole,
 depending on the system parameters.
 If the origin of the luminous supersoft
 X-rays is the atmosphere of a white dwarf that is burning
 hydrogen in shell, it is as hot and luminous as 
post-thermonuclear flash novae, yet no major optical outburst has ever been observed,
 suggesting that the white dwarf is
 very massive (m$\geq$ 1.2 M$_\odot$) and  it is accreting and burning at the high
 rate $\dot m > 10^{-8}$ M$_\odot$ year$^{-1}$ 
expected for type Ia supernovae progenitors.
In this case, the X-ray variability may be due to a very short recurrence time
 of only mildly degenerate thermonuclear flashes.
\end{abstract}


%%%%%%%%%%%%%%%%%%%%%%%%%%%%%%%%%%%%%%%%%%%%%%%%%%

%%%%%%%%%%%%%%%%% BODY OF PAPER %%%%%%%%%%%%%%%%%%

\section{Introduction}
 Very luminous and persistent SSS have been observed since the end
 of the '70ies with the Einstein satellite, but they still pose an unsolved riddle.
60\% or more of  the SSS
 are transient sources,  and we know that the vast majority
of these are post-outburst novae. In novae, the white dwarf (WD) keeps on
 burning hydrogen for a period of time ranging from a week to  years after the outburst,
 with an atmospheric temperature of up to a million K.
 This has been clearly demonstrated in the Galaxy
 \citep[see reviews by][]{orio2012, osborne2015} and in the
 large SSS population of M31
 \citep[][]{orio2006, orio2010, pietsch2005, pietsch2006, henze2013, 
henze2014}.

 However, the nature of numerous SSS  is not yet understood,   even if they
 may hold the key to outstanding
 astrophysical problems. Because of their large intrinsic
 luminosity, these sources are observed in the direction of external
 galaxies, in the Local Group and beyond, up to a distance of 15 Mpc,
in regions of the sky affected by low absorption. 1638 SSS are
 included in the last Chandra source catalog of X-ray sources \citep[][]{Wang2016}.
  It is very likely that
 the majority of SSS are intrinsic in the population
 of the galaxies towards  which they are observed;
 in fact only one SSS out of $\simeq$100 in M31 has
 been found to be an active galactic nucleus in the
 background of the galaxy \citep[][and references
 therein]{orio2006, orio2010}. Some objects in the 
 low luminosity and high hardness-ratio
 end of the SSS in the Local Group are supernova
 remnants, but they do not constitute the majority of the observed SSS. 
 We know now that many SSS that are persistently
 X-ray luminous defy a straightforward classification; this
 is  especially true for the intriguing ones
 observed in galaxies outside the Local-Group, whose luminosity
 appears to be super-Eddington for a star
 of a few solar masses \citep[see][]{liu2011, liu2015}.
  Recently \citet[][]{liu2015} have shown that a very 
luminous SSS, which is persistently supersoft and emitting
 at above-Eddington level for a stellar object,  
 is most likely a micro-quasar hosting a 
 stellar black hole. 

 Many SSS have been proven to be close binaries hosting
 the hottest, most massive accreting and hydrogen burning WDs, 
which   may be on the verge
 of type Ia supernova explosions \citep[SNe Ia; see reviews
 by][]{orio2012, orio2013}. Such WDs represent
 a key to understanding binary evolution and its endpoints.
 Perhaps, by revealing the nature of additional SSS, whether they are 
accreting and burning WDs or not, and by obtaining definite statistics, 
 we will be able to better calibrate SNe Ia for
 cosmological purposes; we may find in fact whether there are different
 types of progenitors, causing deviations from the Phillips
 relationship at low metallicity \citep[see e.g.][]{Meng2011}.
 The models predict that, at very high mass transfer rate $\dot m$,
 the CNO-cycle hydrogen burning on
 the surface of a WD proceeds at such a high rate, that
 all energy is irradiated \citep[e.g.][]{fuji1982, wolf2013}.
 When the most massive, hottest WDs do not
 undergo thermonuclear flashes causing nova outbursts,
 they accrete quietly until either a final explosion
 in a thermonuclear supernova, or a collapse to  neutron star.

 Following \citet[][]{luna2013}, we define symbiotic stars as
 interacting binaries with a red giant,
 asymptotic giant branch star, or exceptionally a supergiant, and
 a compact object of any nature. In the following context we
 will refer to WD-symbiotics as such (as opposed
 to rare symbiotics containing a neutron star or a black hole).
 In the Galaxy, in the Magellanic Clouds and
 in the Draco dwarf spheroidal galaxy, several WD-symbiotics host hydrogen burning WDs
\citep[][and references therein]{orio2013}. 
 However all of them but one, SMC 3,  emit at the low end of the SSS effective temperature
  range, T$_{\rm eff}\leq 200,000$ K, which is characteristic of low
 mass WDs \citep[see][]{starrfield2012,
 wolf2013}.  Since the duration of the residual hydrogen
 burning  phase is inversely dependent on T$_{\rm eff}$
 and WD mass (see Section 3), and low mass WDs may
 have a very long post-outburst residual hydrogen burning phase,
 it is still unclear whether some of these SSS WD-symbiotics are post-thermonuclear
 runaway novae, or whether they are really burning without ever ejecting
 and losing accreted mass.

In contrast with the relatively rich statistics of SSS WD-symbiotics, the census of
persistent SSS  binaries  proven to host a main sequence
 companion and a WD still amounts to only two objects, CAL 83 and SMC 13,
 which have both been monitored for over 30 years. 
 The latter hosts a low mass WD and is not a type Ia supernova candidate
 \citep[][and references therein]{orio2013}, but
 the WD of CAL 83 must be very massive \citep[][]{lanz2005}. There
 is some evidence that another very luminous SSS in M31, 
 Chandra source CXO J004252.5+411539 or r2-12,
 may be a very short period binary \citep[][]{chiosi2014}.

 In this article we present the optical spectrum of yet another
 very luminous and hot SSS in M31, the Chandra source CXO J004318.8+412016 (also
 cataloged as r3-8, as 
 ROSAT source RX J0043.3+4120, and as XMM-Newton source 2XMM
 J004318.8+412017). This source was first detected in 1979 with {\sl Einstein},
 \citep[][]{Trinchieri1991} and has been detected repeatedly in the last
 28 years in many exposures taken with {\sl ROSAT}, {\sl Chandra},
 {\sl Swift}, and {\sl XMM-Newton} \citep[see][and 
 references therein]{orio2010, chiosi2014}. The optical
 spectrum is presented in Section 2. The 
 source  X-ray luminosity undergoes large fluctuations within months;
 we study and discuss the X-ray variability in Section 3.
 Section 4 presents a discussion of the results and we draw conclusions
 in Section 5.

\section{The Gemini spectrum}
 CXO J004318.8+412016  has a an optical counterpart,
 a 22nd magnitude H$\alpha$ emitter \citep[][]{massey2006,
hofmann2013, chiosi2014}. The coordinates of this object in
 the PHAT survey \citep[Panchromatic Hubble
Andromeda Treasury,][]{dalcanton2012} are $\alpha$(2000)=00,43,18.883
 and $\delta$(2000)=+41,20,17.02. This position differs from 0.52$"$ from the
 {\sl Chandra} HRC position determined by \citet[][]{kaaret2002}, 0.38$"$ from the
 {\sl Chandra} ACIS-S position determined by \citet[][]{barnard2014}, 
  0.18$"$ from the {\sl XMM-Newton} 3XMM-DR6  catalog position \citep[][]{rosen2016},
 and 0.23$"$ from the Swift
 coordinates of the 1SXPS catalog \citep[][]{evans2013}.
 In order to evaluate the spatial error box in which we may find
 the optical counterpart, we refer to the online handbooks
 of {\sl Chandra} (the X-ray telescope with the best combination
 of pointing accuracy and spatial resolution) and of the Hubble Space
 Telescope ({\sl HST}).  The {\sl Chandra} absolute astrometry is
 accurate to 0.63$"$ at the 90\% confidence
 level, moreover \citet[][]{kaaret2002} obtains
 an alignment with 2MASS sources with at most
 only a 0.4$"$ discrepancy; the {\sl HST} positions are generally accurate
 within  0.3$"$ within the 90\% confidence level, but the PHAT
astrometry should be even accurate to about 0.2$"$ \citep[][]{dalcanton2012}.  
 At the 90\% confidence level, using the nominal (handbook defined)
 spatial errors boxes of 
 {\sl HST} and the {\sl Chandra} HRC-I (0.63$"$ and 0.3$"$), our
 spatial error box is 0f 0.7$"$ at the  90\% confidence level.

 Several optical
 objects with magnitude between 25 and 27 in the blue F475W filter
 are detected in the PHAT within 0.7$"$; however, as we discuss in detail
 below, we did not find 
 evidence of other emission lines emitters, as expected for
 the optical counterpart of an X-ray binary. The chance of finding an
 emission line star in a 0.7$"$ error box is of course very small,
 probably less than 1\%, so our H$\alpha$ emitter is very likely to be one
 and the same with the X-ray source.
 Moreover, we do not expect a very faint optical counterpart,
 because the luminous X-ray source is either powered by accretion luminosity
 in a binary, or by hydrogen or helium burning,
 which must also be fueled by accretion at high rate \citep[see
 the discussion on accretion luminosity in][]{chiosi2014}. The large soft X-ray flux,
 the extreme softness of the X-ray
 spectrum and the X-ray variability pattern discussed below also suggest
 that the source, most likely, is not a background AGN. To summarize, we suggest
 that there is an extremely high probability that the target of our
 optical observation is one and the same with the X-ray source. 
 
  We observed our target with Gemini-North and the GMOS spectrograph in 
 queue mode (Observing Program GN-2015B-Q-56, PI: J.G.Luna)
 during the nights of 2015/08/20, 2015/08/26 and 2015/09/12. A total of 10 science
 exposures, each with an exposure time of 1650 seconds, 
in long-slit mode, using a 0.75$"$ slit and the B600 grating centered
 on 5600/5650 \AA, were obtained and used for the data analysis presented here.
 The observing conditions during the observations were photometric,
 with seeing $<0.75"$ and dark skies. The spectrum was
 binned with a 2x2 binning in both spectral and
 spatial direction, and the spectral resolving power
 was R=1700.  The resulting median spectrum is shown in Fig. 1.
 
The data reduction and spectra extraction were done using a custom semi-automatic
 python script \footnote{available from http://github.com/rkotulla/gmos-longslit}
 following the standard Gemini/GMOS data reduction script.
 Each frame was corrected for bias and dark-current using the archive-provided master
 calibration products, and flat-fielded with the flat-field frames taken along 
with the science frames. The wavelength calibration was performed by
 means of Cu-Ar arc lamp spectra
 taken after each set of science frames, with spectrograph settings identical to that of
 the science frames. Each frame was also cleaned of cosmic-ray hits using the appropriate
 setting in the \texttt{gsreduce} IRAF task. Each reduced longslit spectrum was then
 rectified  compensate for spectral curvature using the wavelength solution derived from the 
arc-spectra. Absolute flux-calibration was done by calibrating the spectral response 
function using a spectrophotometric standard star (G191B2B) observed as part of the 
observing project. Sky-subtraction is without doubt the largest uncertainty 
given the large number of unrelated sources in the vicinity (see Fig. 2), 
preventing us from isolating actual sky from background emissions. To isolate 
night-sky emission lines we applied a median-filter along the spectral direction to 
isolate the smooth continuum from emission lines, and subtracted the latter from the 
rectified long-slit spectra, providing us with a line-free spectrum, but with night-sky 
and background continuum still intact. To extract the one-dimensional spectrum and 
finalize the sky-subtraction, we integrated the spectrum over the spatial extent of 
the H-alpha emission line, and subtracted the scaled median of the direct 
vicinity along either side of the slit. The final spectrum, presented in 
Fig. 1, was then computed as the mean spectrum of all ten individual spectra, 
rejecting outliers (e.g. remaining cosmics) via an iterative sigma-clipping algorithm. 

We note that across the entire region covered by the slit we find weak H$\alpha$ and 
[N II] emission, and cannot rule out a low-level contamination of the extracted spectra 
from unassociated background emission, in particular in the case of the weak [N II]
detection. The slit was oriented with the parallactic angle, so 
 the single exposures were taken at different hour angles, and in each
 of them
 the slit had a different orientation with respect to the sky coordinates.
 The emission lines we detected and measured were the same, and had the same
 characteristics,  
 in each of the single exposures, so we are confident that 
 the background and neighborhood contamination is negligible, 
 apart from the low flux level diffuse H$\alpha$ and [N II] emission
 mentioned above.
 The emission spectrum we present here originates 
 in our target star, the only one that was always in the slit. 

 This spectrum is characterized by narrow, strong emission lines
of the Balmer series, several He I lines and a relatively weak (compared
 to what we usually see in the hottest
 WD-symbiotics) He II line at 4686 \AA. Lines due to very high excitation
 or ionization stages, like coronal lines, are missing in this source.

 The measurements
 of the flux in the lines for the rest and measured wavelengths are shown in Table 1. 
 The lines are blue-shifted by -349.8$\pm$23.4 km s$^{-1}$, which 
 is consistent with an object intrinsic in M 31. The systemic
 velocity of the galaxy is $\simeq$-295 km  s$^{-1}$ \citep[][]{Drout2009,McConn2012}. In the heliocentric velocity field measured by \citet[][]{Emerson1976}
 we find that the  expected velocity is -240$\pm$30 km s$^{-1}$ at the star's location. 
%while following the parameterization given
% by \citet[][]{Drout2009, Evans2015} 
% we expect about -185  km  s$^{-1}$.
 \citet[][]{Evans2015} show that a
 difference of -90  km  s$^{-1}$ from the expected velocity is above
 the average, but it is not unusual at all for M31 red giants.
 We conclude that our object is in the thick disk or halo.
 We also note that foreground objects have positive velocity difference from
 the expected one, being 
blue-shifted by less than 150 km s$^{-1}$, so the velocity we measured proves M31
 membership. 

The emission lines are well detected in all the
 single exposures and in the median spectrum in Fig. 1, and
we found no indication of clear variability of any of the lines' flux
 or line centers between different exposures.

 Although the prominent Balmer lines in emission and the lines of neutral 
 helium are typical of both Be stars and WD-symbiotics, 
 our initial classification of the secondary as a B[e]  star
 \citep[][]{orio2015}, 
 due to the tentative identification of [Fe II] lines that
 are typical only of B[e] type stars, could not confirmed,
 in fact we found that those lines are not detected at a statistically
 significant level in the stacked spectrum.

 The strong Balmer decrement
 (H$\alpha$/H$\beta$=6.7 and H$\gamma$/H$\beta$=0.26) seems to
 imply high reddening; it would translate in fact into 
 E(B-V)=0.71 (or A$_{\rm V}\simeq$2.2 mag for a Galactic-type extinction law).
We suggest that the Balmer decrement is
 due to high optical depth in the binary.
 The high extinction implies N(H)$\simeq 3.55 \times 10^{21}$ cm$^{-2}$
\citep[following][]{Burstein1982}, or N(H)$\simeq 3.94 \times 10^{21}$ cm$^{-2}$
\citep[following][]{Predehl1995},  higher values than the best fit
 value in most X-ray observations and only marginally consistent with  
 most X-ray spectra, although a 2015 exposure close to
 the Gemini observation indicates that N(H) may have increased 
 with respect to previous observations (see Section 4). 
 
In the direction of M31 the interstellar absorption
 is very low, E(B-V)$\leq$0.07 
 (N(H)$\simeq 3.50 \times 10^{20}$ cm$^{-2}$ \citep[following][]{Burstein1982}
 or $\simeq 4.06 \times 10^{20}$ cm$^{-2}$ \citep[according to][]{Predehl1995})
 however the column density of neutral hydrogen
 inside M31 varies by a large factor. In \citet[][]{orio2010} we found a dust
 lane in the region of the source, and suggested that the SSS must be located in front
 of it, or else it would not have been detected. However, this
 is not necessarily so, considering accurate
 and spatially resolved data on dust extinction, recently
 published by \citet[][]{dalcanton2015}. 
 At the position of the source, in the maps
 of these authors we find A$_{\rm V}\simeq$0.6,
 corresponding to
 E(B-V)$\simeq$0.2 and about N(H)=10$^{21}$ cm$^{-2}$ for a Galactic-type extinction
 law (however, the map shows large patchiness on very small scales, and
 there is much higher extinction in the zones neighboring the source).
  \citet[][]{orio2006}  found also
 that the column density is likely to be variable, 
 and that the  2 $\sigma$ lower
 limit in different observations
 is N(H)$\geq 9 \times 10^{20}$ cm$^{-2}$
(which is consistent with  A$_{\rm V}\approx$0.6).
 The variability may be due to an unstable 
 wind, causing changing intrinsic absorption within the binary system.
  \citet[][]{orio2006} also noted that the column
 density derived from the best fit increases, by up to a factor of 5, when
 the X-ray source is at maximum luminosity (thus the repeated X-ray dimming
 is not likely to be due to increased column density).

 The emission line of He II at 4686 \AA \
 is  often observed in WD-symbiotics \citep[][]{luna2005, miko2017}, and it is
 typical of supersoft X-ray sources. 
 This line is often detected in many hot,
 accreting binaries, and it is always present in the
 spectra of accreting WDs.  Because
 it is produced with a high ionization potential, it needs a 
 hot environment and it usually originates
 near the WD. The flux in this line corresponds to a luminosity
 3.7 $\times 10^{33}$ erg s$^{-1}$ for the  M31 distance, which is consistent
 with an accretion disk illuminated by ionizing radiation,
 either coming from the very 
massive hydrogen burning white dwarf, or from the disk itself if it 
surrounds a stellar mass black hole undergoing supercritical accretion
 with optically thick outflows. It is puzzling, however, that the
 ratio of the intensity of the He II and H$\beta$ line is only 0.39,
 while usually in the other with hydrogen burning WDs, 
 this ratio is about 1 \citep[see e.g. SMC
 3, AG Dra and Lin 358][]{orio2007, Munari2002}. 
Also the ratio of the He I lines relative to H$\beta$ is
 unusually large (He I $\lambda$5875/H$\beta \simeq$0.7). 

Because we cannot measure the continuum in the optical spectrum, 
 in the next Section we use archival data to analyse the
 nature of the secondary. Since we will show that it appears to be a red giant,
 we note here an interesting fact that constrains the
 spectral type: the absence of a TiO band, a feature that should have been
 measurable, suggests a classification
 of the red star as a giant of spectral type K or earlier.
 
We must also note the absence of the Raman scattering O VI line
 at 6825 \AA, and of  strong coronal lines of [Fe X],
 which were observed in the spectra of SMC 3 and other
 hydrogen burning WD-symbiotics \citep[see][]{orio2007, miko2014}.
  With a central source at a temperature
 close to a million K (see Section 4),
 the [Fe X] coronal line should have been produced in the
 optical spectrum unless the symbiotic nebular medium was much denser than in
 most symbiotic  and  the spectrum arises in a nebula
 with electron density above the critical one for this line, n$_e \simeq 5
 \times 10^9$ cm$^{-3}$ \citep[][]{nagao2002}.  

\begin{figure*}
	\includegraphics[width=120mm]{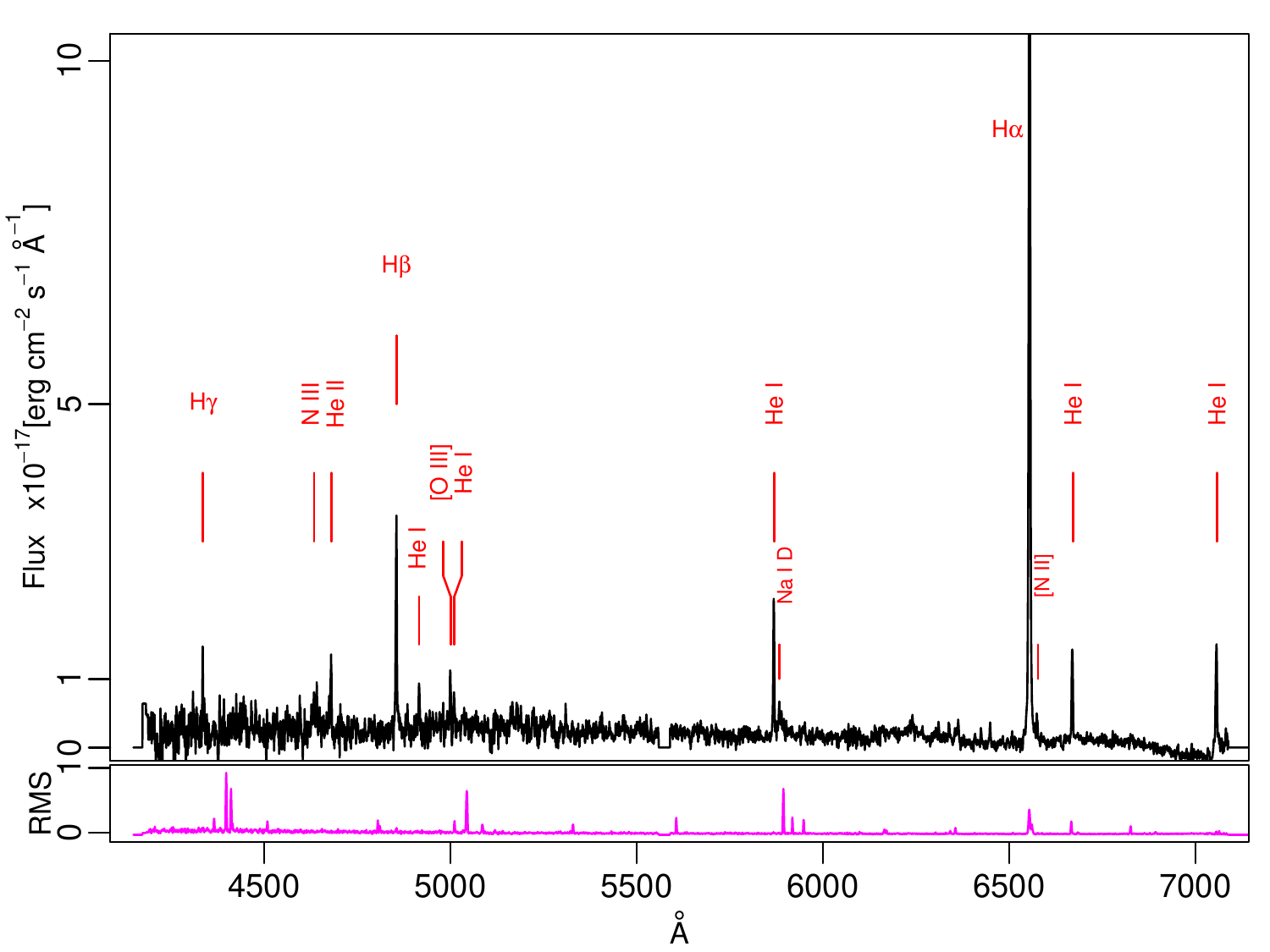}
%\vspace{-4.2cm}
    \caption{Median spectrum obtained with a stack of 10 
 exposures of the SSS binary CXO J004318.8+412016 in M31,
 obtained with the Gemini North telescope and the GMOS spectrograph.
  The purple line shows
 the root mean square error, indicating the deviation from the median of
 the stacked spectra. the peak of the H$\alpha$ line is close to 14 $\times 
 10^{-17}$ erg cm$^{-2}$ s$^{-1}$
\AA$^{-1}$, but we have cut the y-axis to allow seeing more detail in the other lines.}
\end{figure*}
\begin{figure}
\includegraphics[width=80mm]{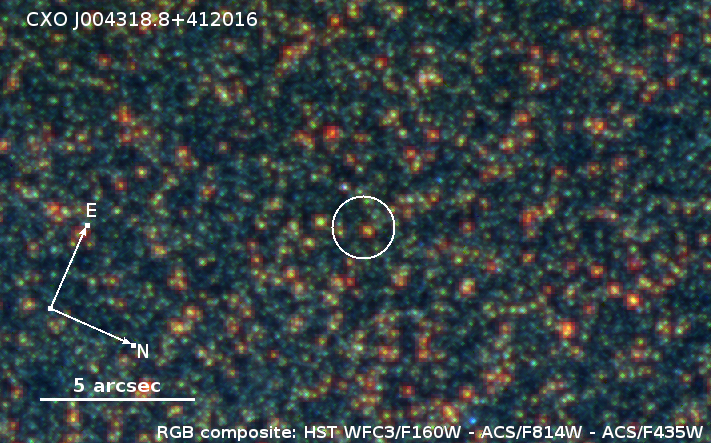}
\caption{Composite figure of images in 3 filters of our target's field in the PHAT. 
 Our optical target is the red and most luminous in 
 the circle, which has a 1$''$ diameter.}
\end{figure}
\begin{table}
	\centering
	\caption{Emission lines in the optical spectrum of r3-8,
 and their flux, when measurable.}
	\label{tab:flux_table}
	\begin{tabular}{lccr} %  columns, alignment for each
		\hline
 Line & Rest $\lambda$ & Measured $\lambda$  & Flux $\times 10^{-17}$ \\
      & (\AA)  & (\AA) & (erg s$^{-1}$ cm$^{-2}$) \\
		\hline
 H $\gamma$ & 4340.46 & 4335.84  & 2.83$\pm$0.01  \\
 N III      & 4640.64 & 4634.7 & \\
 He II      & 4685.91 & 4679.99  & 4.21$\pm$0.11  \\
 H$\beta$   & 4861.33 & 4855.55  & 10.89$\pm$0.03  \\  
 He I       & 4921.93 & 4916.76 & 3.83$\pm$0.67 \\
 O [III]    & 5006.84 & 5000.51  & 3.09$\pm$1.02   \\
 He I       & 5015.68 & 5010.55  & 2.49$\pm$0.05 \\
 He I       & 5875.62 & 5868.69 & 7.31$\pm$0.03 \\
 Na I D      & 5889.95 & 5883.42 & 1.54$\pm$0.25 \\
 H$\alpha$  & 6562.80 & 6555.05 & 72.99$\pm$0.16 \\  
 N [II]       & 6583.46 & 6575.31 & 1.98$\pm$0.03 \\
 He I      &  6678.15 & 6669.85 & 6.18$\pm$0.14 \\
 He I       & 7065.71 & 7056.69 & 10.35$\pm$0.04 \\
\hline
	\end{tabular}
\end{table}
\begin{table*}
\centering
\caption{ PHAT and LGS average magnitudes of the optical counterpart,
 as measured, and converted to absolute magnitude and dereddened 
 assuming  E(B-V)=0.20 (fourth column,   mag${_a1}$ ) and E(B-V)=0.75 (fifth
 column,   mag${_a2}$ ).  The sixth column reports the actual
 measured magnitudes in the catalogs, and the LGS ones are accompanied by
 the average statistical error (seventh column) for the given filter;
 however this is a lower limit for a crowded field. The statistical errors for the PHAT
 measurements are around 0.05 mag in relatively 
 crowded fields like this one \citep[][]{williams2014}.  }
\label{tab:mag}
\begin{tabular}{lccccccr}
\hline
 Filter & Center $\lambda$  & Bandpass  & mag$_{a1}$ &  mag$_{a2}$ & mag & $\Delta$mag \\
        &         (\AA)     & (\AA)     &          &           &     & \\ 
\hline
 F275W & 2710 &  164.5 &  -3.372 & -6.882 & 21.948 & \\
 F336W & 3355 &  158.4 &  -4.53  & -7.110 & 20.930 & \\
  U    & 3650   &  660    &  -3.376 & -6.195 & 21.685 & 0.027 \\
  B    & 4450   &  940    &  -2.127 & -4.187 & 23.133 & 0.10 \\
 F475W & 4774 &  421.2 &  -2.533 & -4.423 & 22.657 & \\
  V    & 5510   &  880    &  -2.527 & -4.207 & 22.433 & 0.073 \\
  R    & 6580   &  1380   &  -3.017 & -4.297 & 21.943 & 0.071 \\
  I    & 8060   &  1490   &  -3.819 & -4.747 & 21.003 & 0.042\\
 F814W & 8030 &  663.3 &  -3.807 & -4.759 & 21.001 & \\
 F110W & 11534   & 1427   &  -5.082 & -5.582 & 19.578 & \\
 F160W & 15369  &  1341   &  -5.838 & -6.168 & 18.732 & \\
\hline
\end{tabular}
\end{table*}
\begin{figure*}
\includegraphics[width=0.47\linewidth]{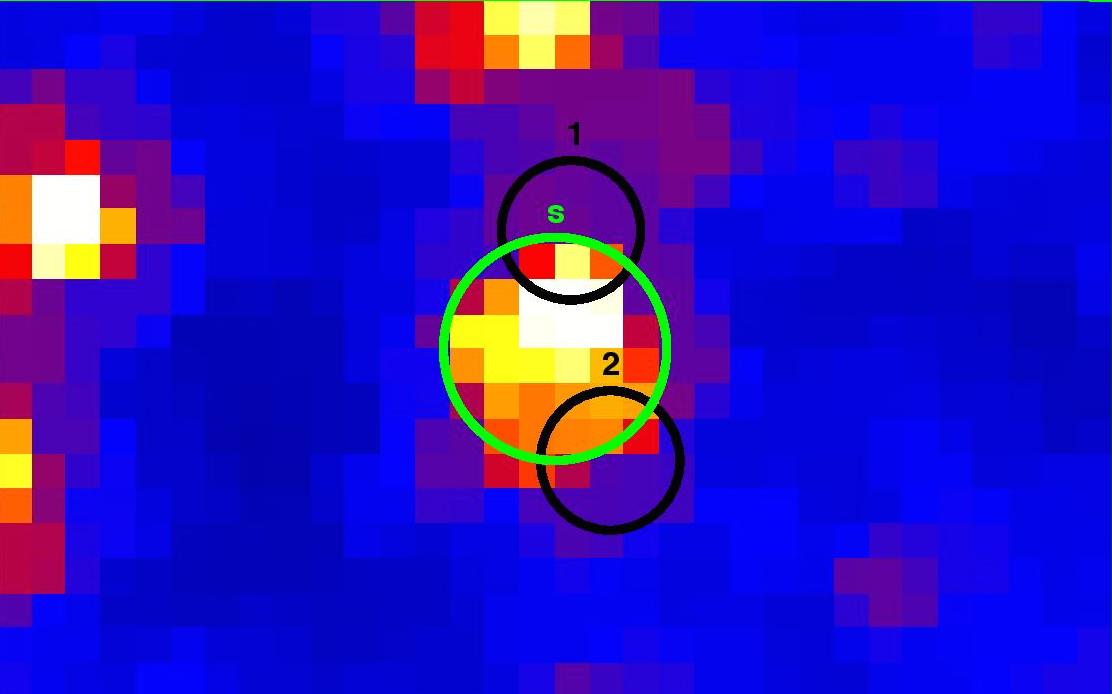}
\includegraphics[width=0.47\linewidth]{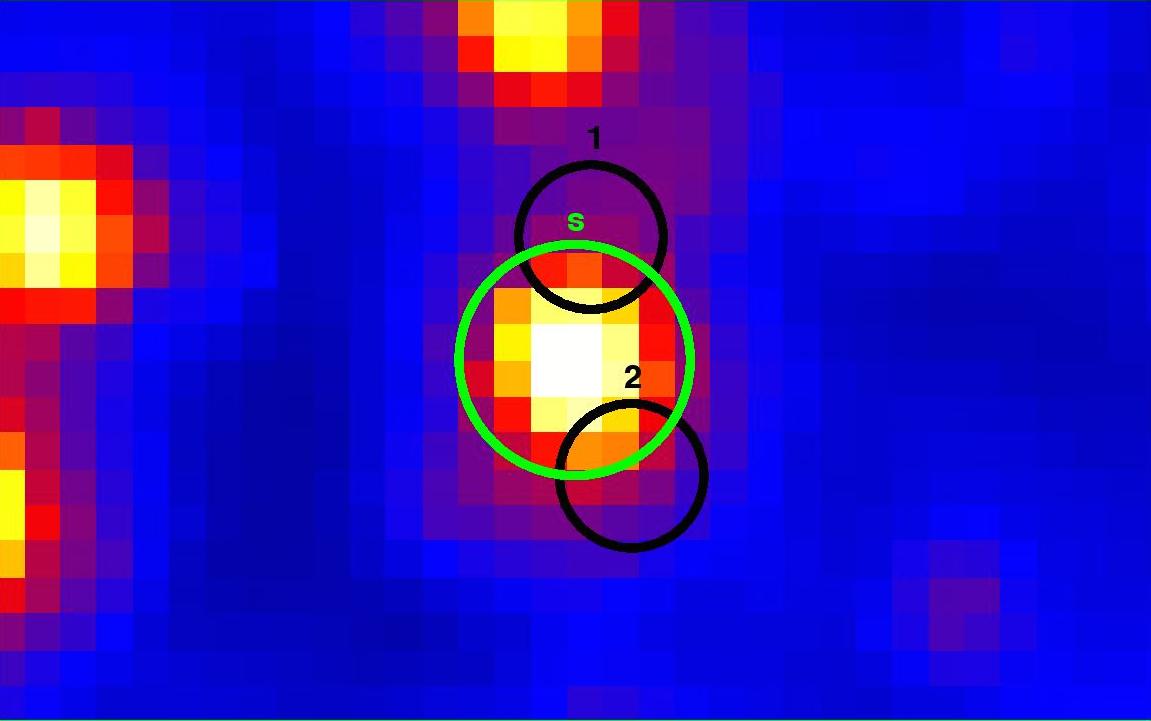}
\includegraphics[width=0.47\linewidth]{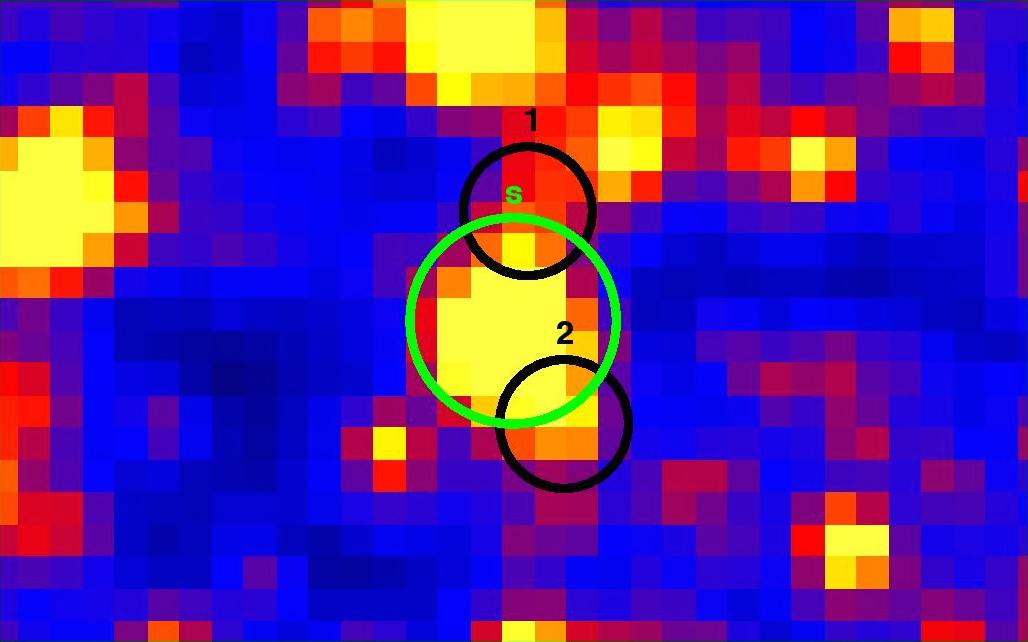}
\includegraphics[width=0.47\linewidth]{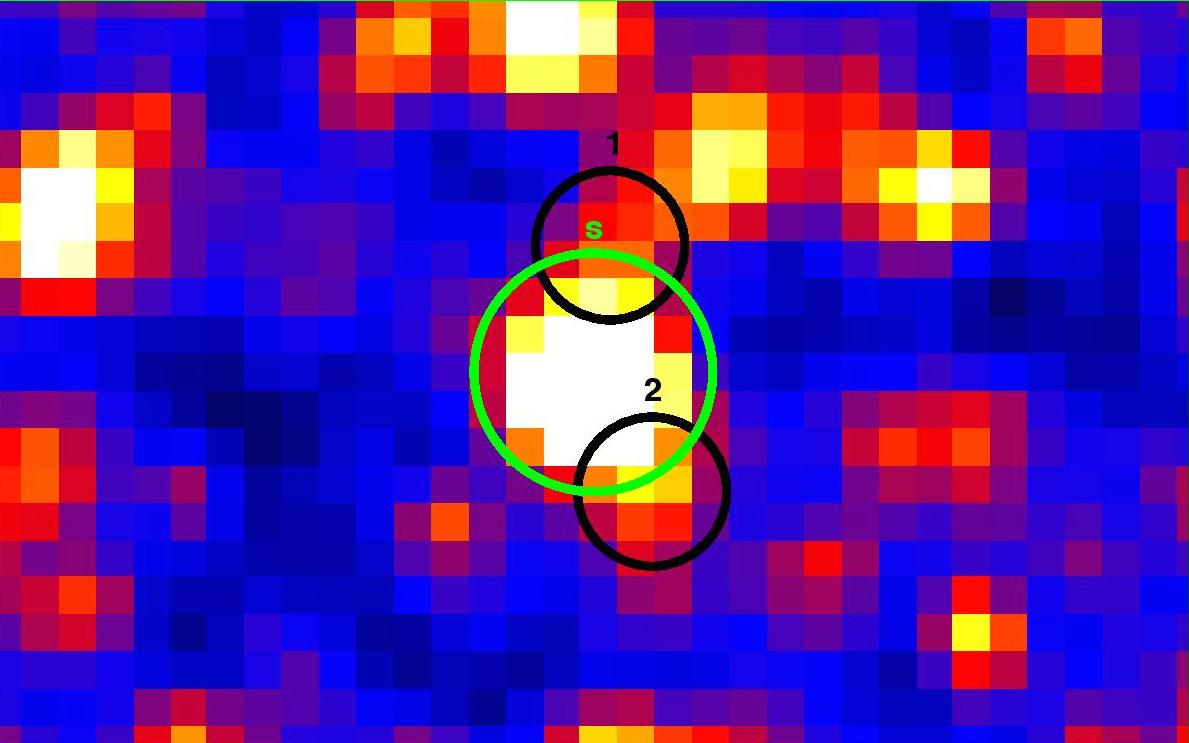}
\includegraphics[width=0.47\linewidth]{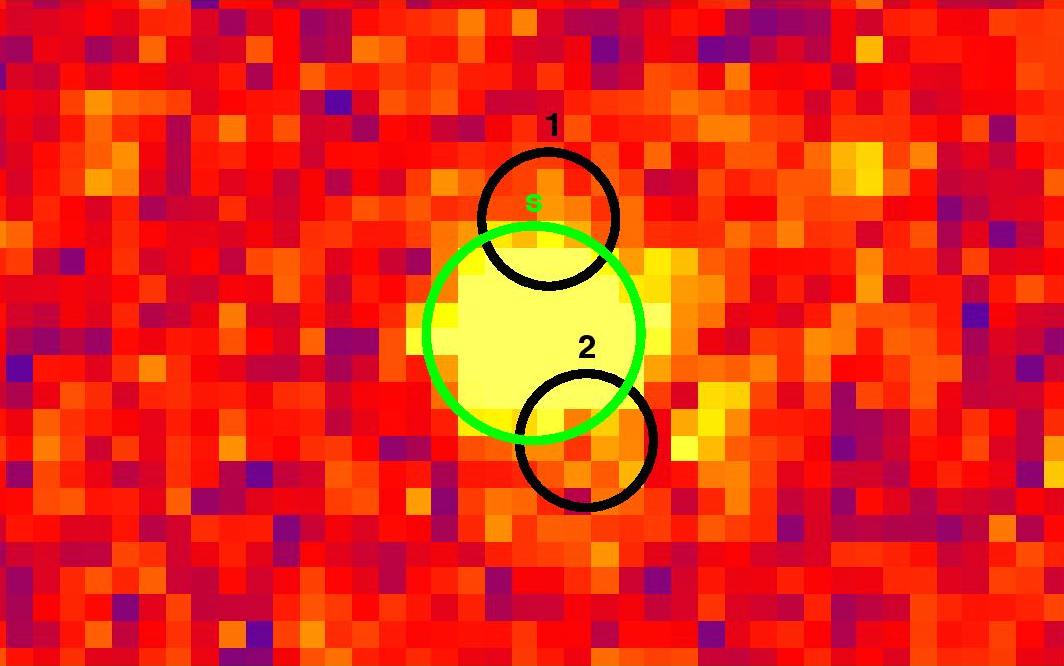}
\includegraphics[width=0.47\linewidth]{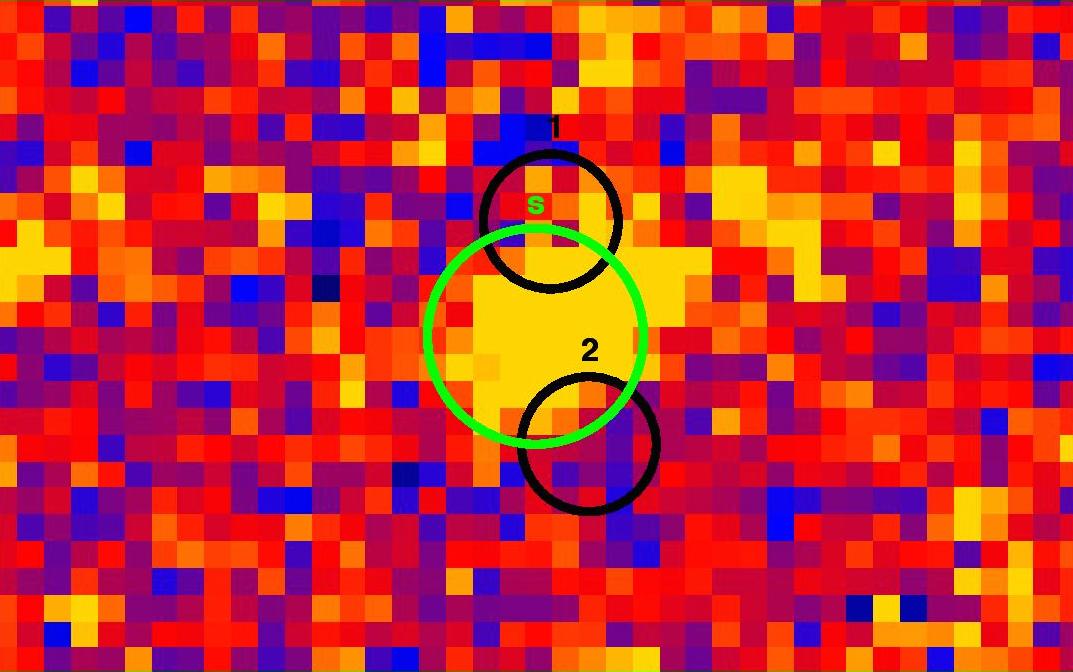}
%\vspace{-1.0cm}
\caption{The field of CXO J004318.8+412016 
 around the star whose spectrum we are presenting, in the Brick 3 field
 of the PHAT, with the F110W filter (top left), F160W(top right), F814W (middle left),
F475W (middle right), F336W (bottom left) and F275W (bottom right). Stars 1 and 2
 are marked by a circle of 0.09$"$ radius, and represent the two objects
 with overlapping wings of the PSF, measured in the PHAT (they
 are hardly detected in the IR and below threshold
 limits in the UV). Our target is in a 0.15$"$ radius green circle. The
 images are oriented with North on top, the field has dimensions of 
 1.6$"$x1$"$. }
\end{figure*}
\begin{figure*}
\includegraphics[width=0.47\linewidth]{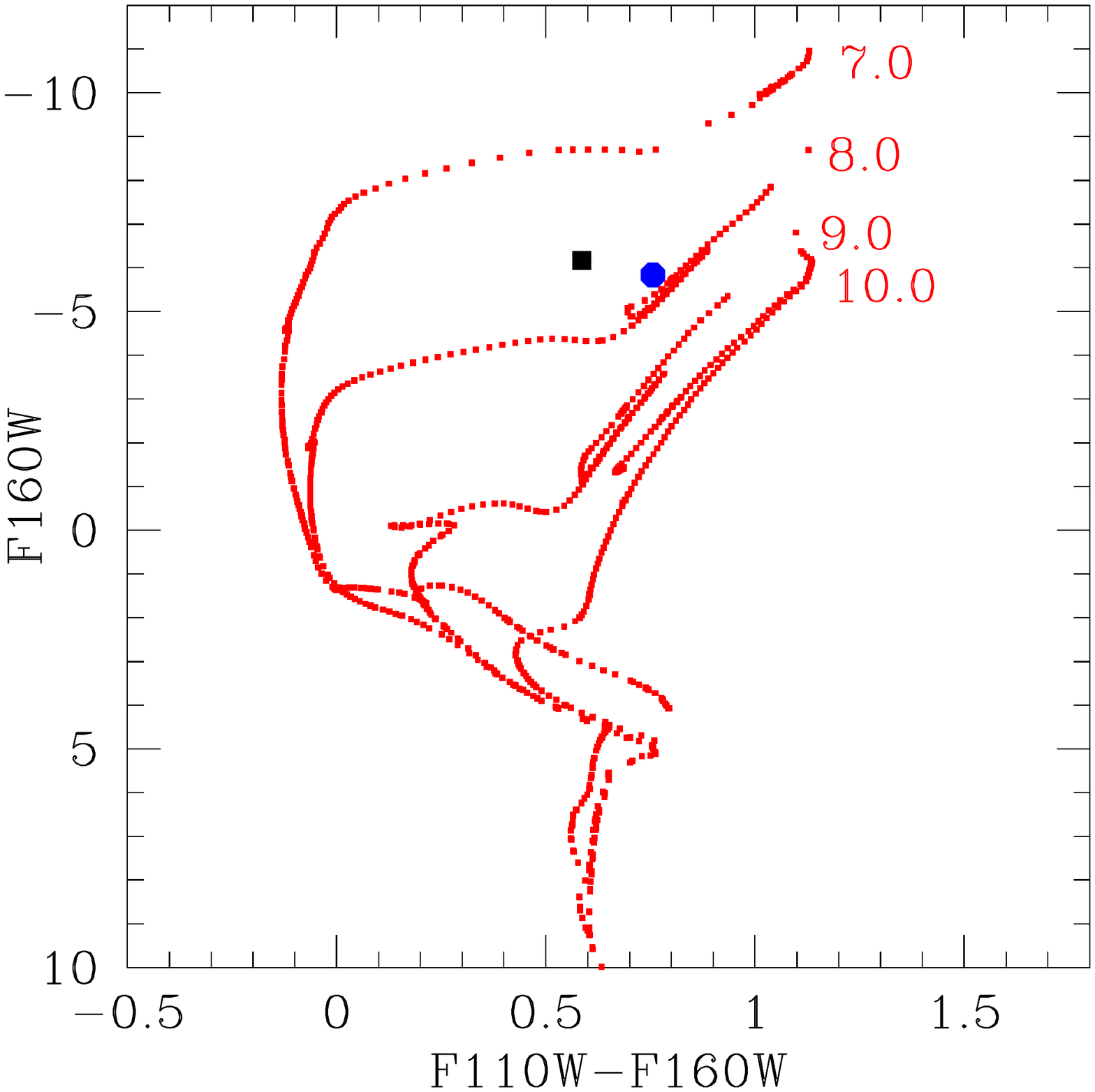}
\includegraphics[width=0.47\linewidth]{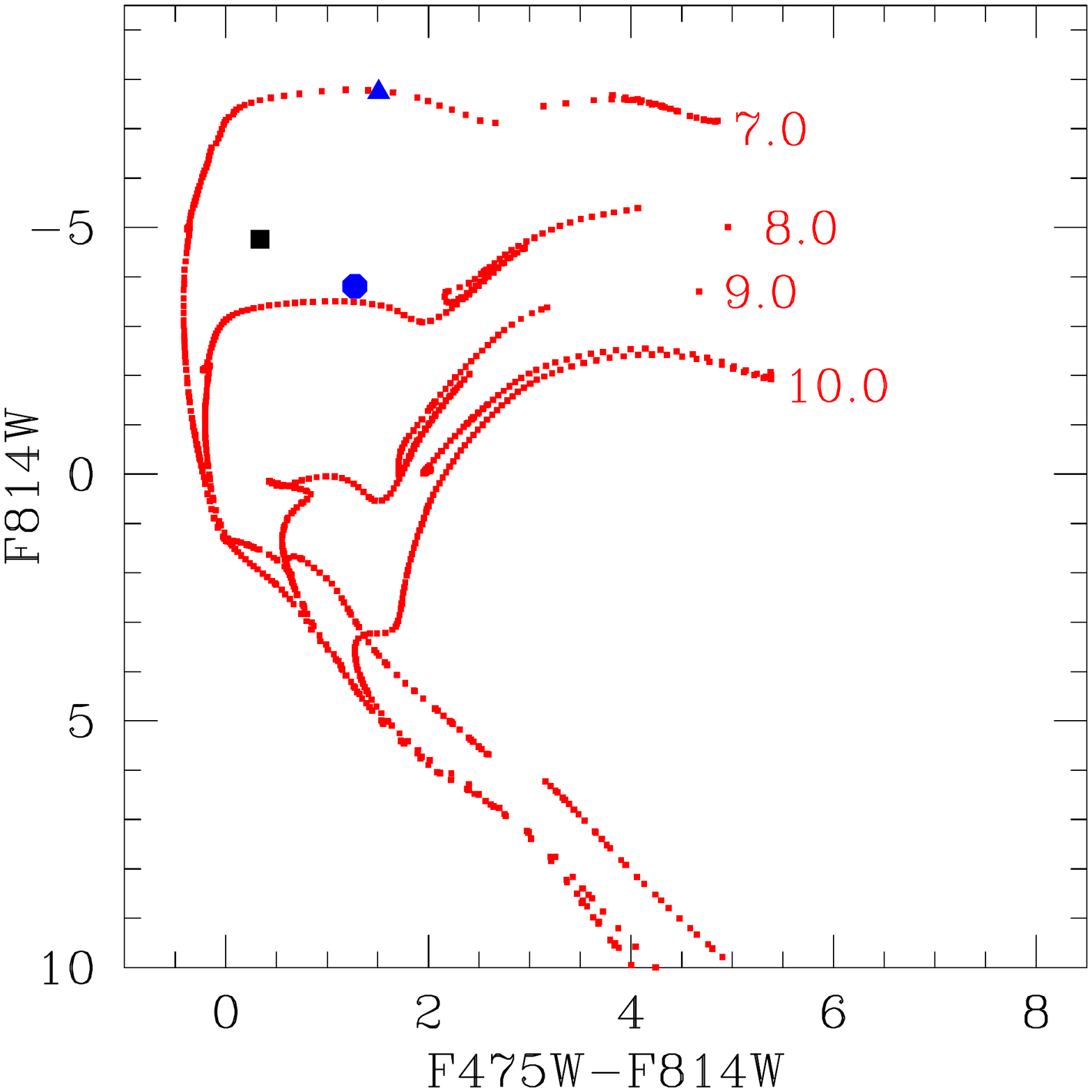}
\includegraphics[width=0.47\linewidth]{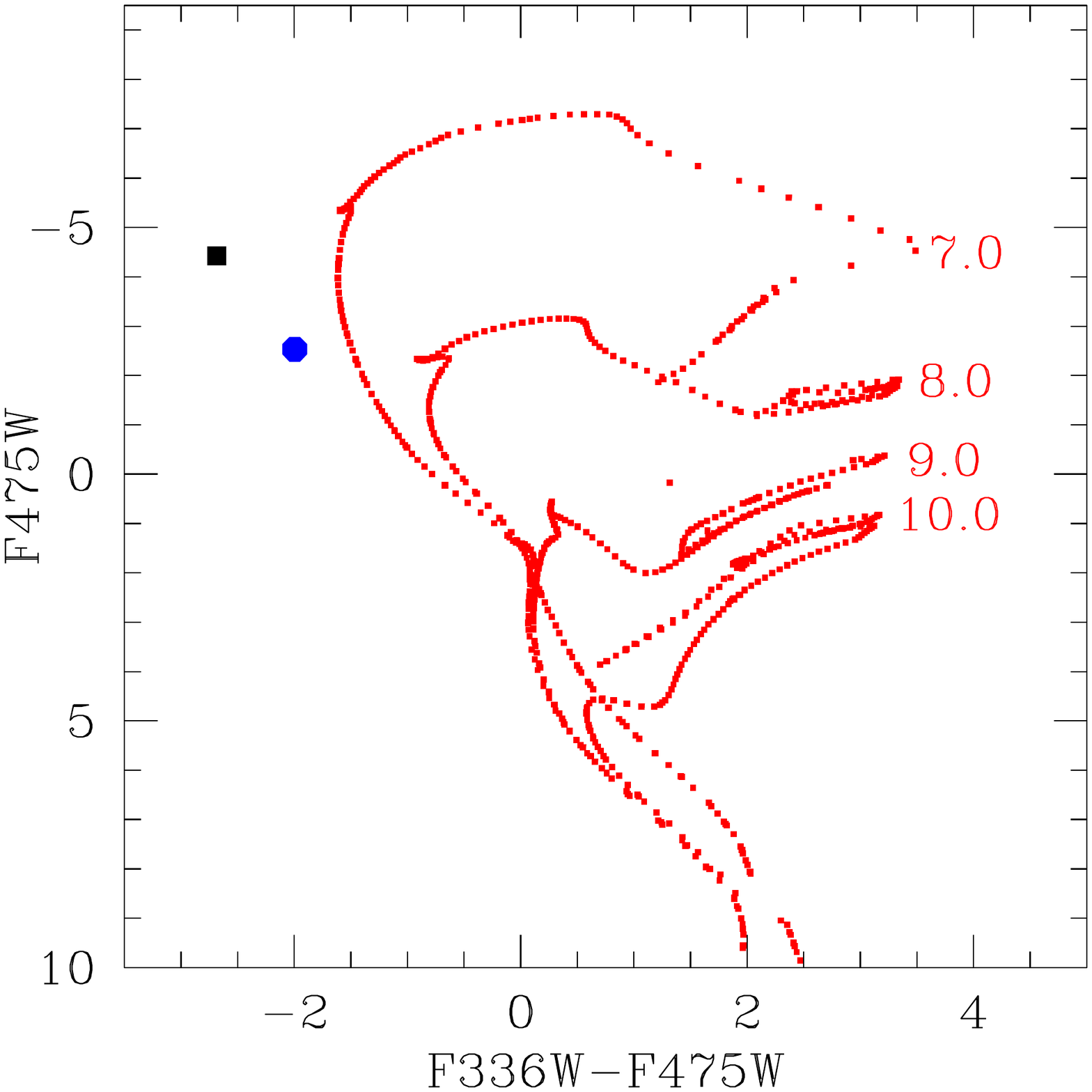}
\includegraphics[width=0.47\linewidth]{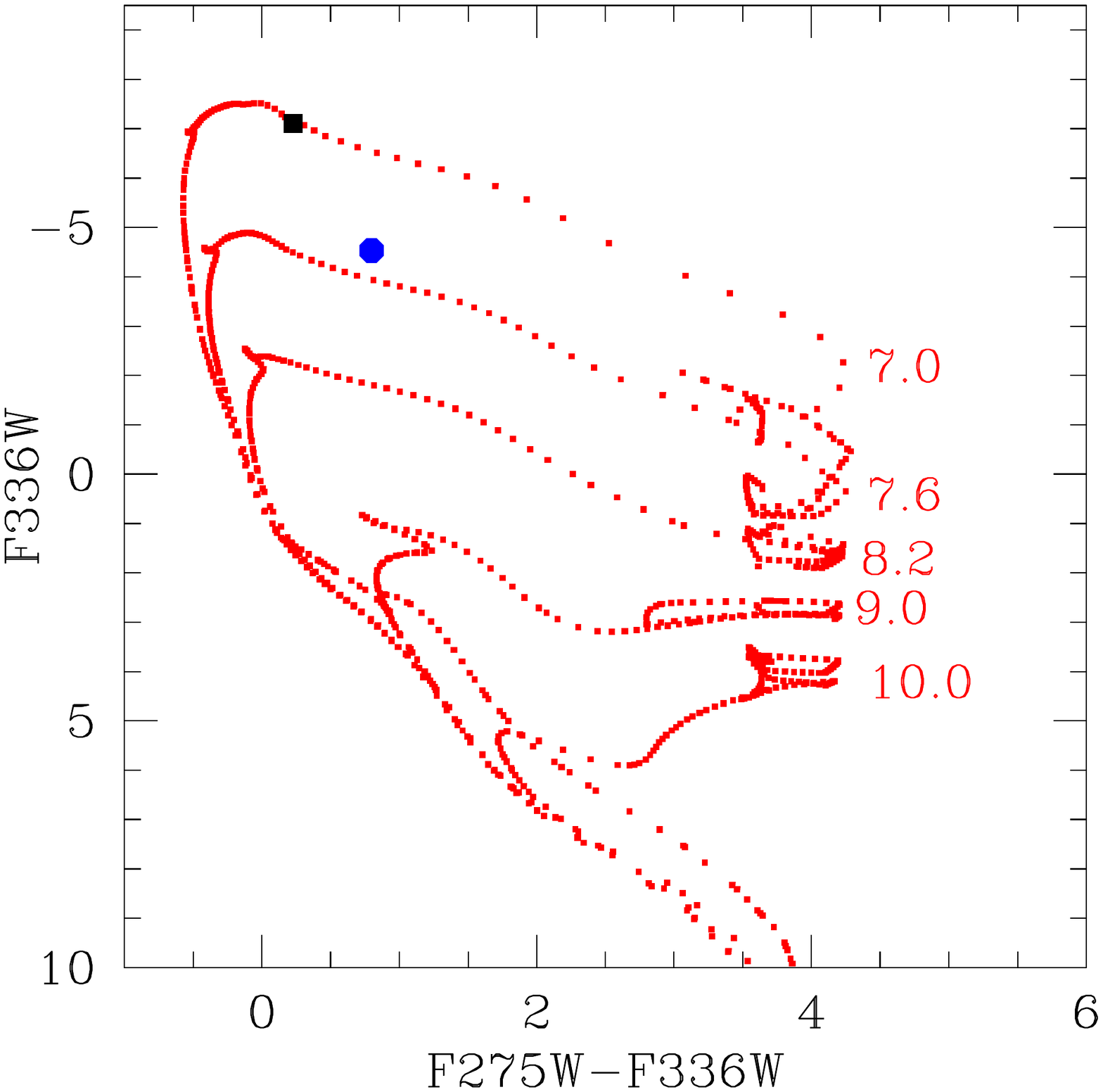}
\vspace{-1.0cm}
\caption{Color magnitude diagrams with isochrones of \citet[][]{Bressan2012} and position
 of CXO J004318.8+412016. We assumed E(B-V)=0.20
  (blue circles) and E(B-V)=0.71 (black squares) in dereddening the 
 magnitudes. The triangle in the upper right panel indicates
 the position of SMC 3 in the diagram. The units are magnitudes on
 the x and y axis.}
\end{figure*}
\begin{figure*}
\includegraphics[width=80mm]{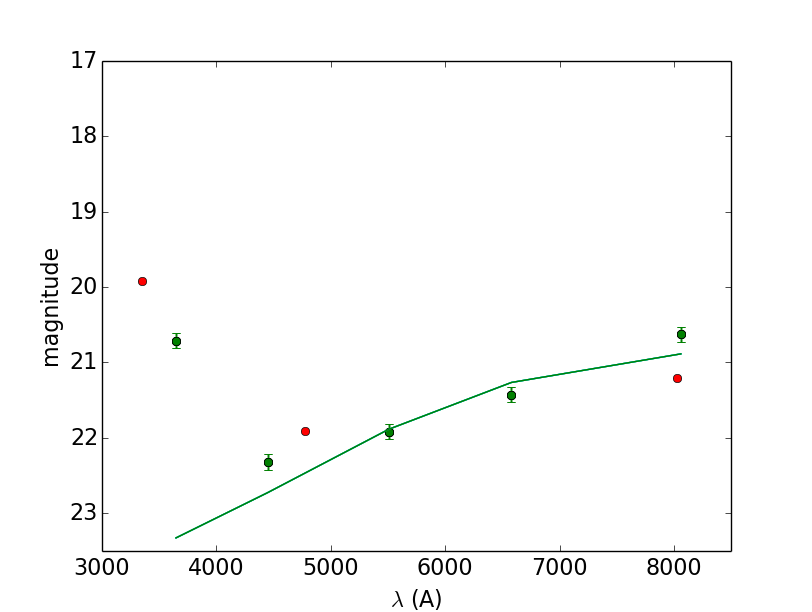}
\includegraphics[width=80mm]{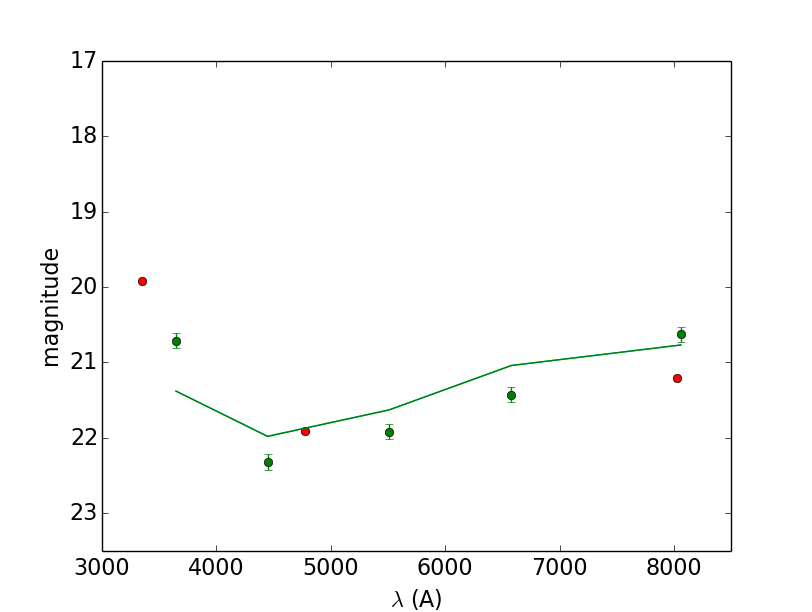}
\includegraphics[width=80mm]{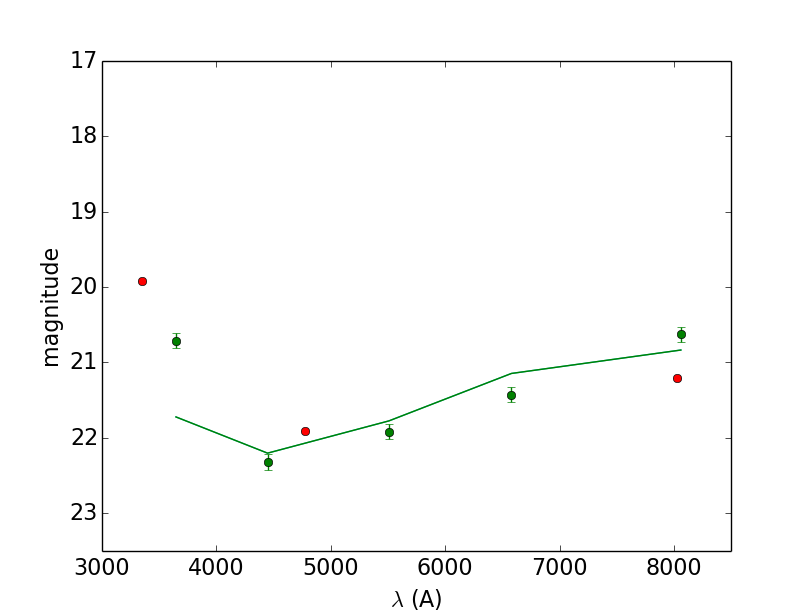}
\includegraphics[width=80mm]{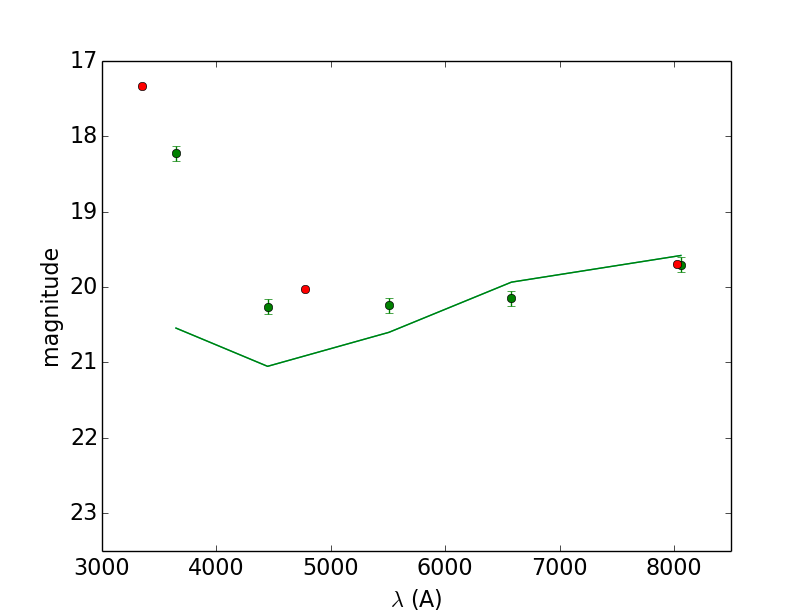}
\caption{ The green solid lines show the 
fit to the measured catalogs' magnitudes with a red giant   (top left panel). The
 other panels show the fits by varying the red
 giant temperature and including 
 an  accretion disk: around a WD of $1.3 M_\odot$
 in the top right and bottom right panels with E(B-V)=0.20 and E(B-V)=0.71,
 respectively; around a black hole of 5 $M_\odot$ and  E(B-V)=0.20  in the
 bottom left panel (see text for
 details). The LGS magnitude are indicated by the green dots, the PHAT ones by
 the red dots.
}
\end{figure*}
\section{The nature of the secondary and the fit with an accretion
 disk model}
Because we cannot reliably measure the continuum in our spectra
 due to the elevated sky background, we resorted to the photometric
 catalogs to examine the spectral energy distribution (SED) of our 
target from infrared (IR) to ultraviolet (UV).
In Table 2 we give the magnitudes from the PHAT survey \citep[][]{dalcanton2012, williams2014, johnson2015}
 as in the most recent version of the data release; these measurements
 supersede the values given in \citet[][]{chiosi2014}, because
 an initial analysis of the Brick containing this object has been
 completed and revised by the PHAT team, see \citet[][]{williams2014}.
 Additional measurements, albeit with a larger error, were obtained in 
 the LGS (Local Group Survey, Massey et al. 2006) by stacking exposures
 taken in the course of over a year.
 The error bars in Table 2 are the mean errors of the LGS final photometry,
  but this field is sufficiently crowded to cause quite larger photometric errors.
 In Fig. 3 we show the PHAT
 coadded images of Brick 3, containing CXO J004318.8+412016, in
 the six PHAT filters, corresponding
 to the UV, optical ultraviolet, blue, optical infrared and two 
 IR bands (see Table 2).

 There is a marginal, partial overlap of our target, encircled in
 green in the Figure and marked with ``s'' , with stars
 no. 1 and no. 2 in the optical and IR filters. However, the PHAT indicates 
 a quality flag of ``reliable'' for the magnitudes measured for
 all the three objects; moreover star 2 is measured to be at
 least 2 magnitudes fainter than star no. 1 in all filters, and star 2 is
 about 3 magnitudes fainter. We conclude that the crowding should
 not have affected the measurement in a very significant way, although the
 error may be larger than the average value in the PHAT. 

 Orbital modulations or  other type of variability may make the photometric
 measurements less significant to derive
 the spectral energy distribution (SED), because some of the PHAT images were 
 obtained at different epochs. Our target was observed in
 field 9 of Brick 3  of the PHAT on 2013/7/15 in the F275W, 
 F336W and f160W filters, on 2012/12/9 in the F110W filter, and 
 in fields 8 and 9 on 2012/6/30 and 2012/7/1 respectively, in
 the F814W and F475W filters. In the LGS, for each filter
 the exposures were repeated on different dates in 2001
 September and November, and 2002 December. The photometry of
 this target can be done with a small error only by stacking the LGS images,
 so the catalog magnitudes in the Johnson filters are the average of 4-6 exposures.    
  Orbital modulations are expected to occur on time scales of the
 order of a year for a symbiotic, and of the order of several weeks for
 a Be binary. We examined the single LGS exposures in the single deep images
 obtained with the U filter and 
 measured relative magnitudes, concluding that, even in exposures taken
 after about a year, there is no variability larger than 0.1 mag.
 The same is true for the two optical filters of the PHAT in which
 the exposure was repeated after one day. The several
 repeated LGS V and B exposures
 are quite shallow for this target, but we also examined them, finding no evidence
 of large variability. We do note that there is a difference of almost
 one magnitude between the two close wavelength ranges, that of filter U of the LGS and
 of filter F336W of the PHAT, so there may have been a large variation on a timescale
 of 10 years. 
 However, it is likely that the color indexes are relatively constant in each 
 catalog, based on images obtained within little over a year, 
 within possible fluctuations of $\approx$0.15 mag.
 We note that in \citet[][]{orio2006} a WIYN telescope 
 image was presented, in which our target was not detected with the blue filter, and  
 an upper limit of B$<$23.5 was claimed, which would imply a
 variability amplitude of at least 0.36 mag. However, we checked the data again
 and found a typo in the caption; the upper limit 
 for the detection was B$\simeq$22.5 and the detected
 and measured star marked in the figure has B$\simeq$21.7, not B$\simeq$22.7.
 
 In Fig. 4 we use the average PHAT magnitudes and color
 indexes to show the position
 of the optical and UV object in the color-magnitude diagrams.
 We examine two hypothesis, corresponding to two
 extreme values of E(B-V), E(B-V)=0.20 indicated by the PHAT
 at the location of the source, and the value derived from Balmer
 decrement in the 2015 Gemini spectrum,
 E(B-V)=0.71. We remind that the range of column density
 evaluated from the (albeit low S/N) X-ray spectra is consistent with
 these two extremes, and it is likely to be variable.
 This is discussed more in Section 4.
 
CXO J004318.8+412016 has the colors of a young red giant, but there is an excess
 in the magnitudes measured with the Johnson U and F336W filter,
whose bandpass is close to that of the U.  
 First, we compared our object with 
 evolutionary tracks of populations of different ages. The infrared
 and optical colors
 of the optical counterpart are consistent with those a red giant of 
about 100 million years for the lower reddening
 value and the IR colors would indicate the helium burning loop,
which is a short lived phase, but not impossible to detect. The higher value
 of the reddening would imply a younger age of our source. 
   The red component of SMC 3 is more luminous and consistent with
 an age of only 10 million years. 

 However, when we examine the U and UV colors we find 
 that the HST filter close to the U band places our target 
 on the left of the evolutionary tracks (F336W-F475W).
 In the U filter
 there is no significant contamination of other nearby objects
 (stars 1 and 2, and other objects in the nearby fields, are not
 U-bright, as Fig. 3 clearly shows). 
The reddening free value Q=(U-B)-0.72(B-V), which is negative and around -1
 for B star, 
 has a very large value of $\simeq$-1.9 in the LGS, which
 is very unusual. We note that the F275W-F336W color index is still consistent 
 with a giant, although with a more luminous one, and of younger age
 (so young that
 seems to have been ruled out by the kinematics,
 which, as we mentioned, point at the thick disk or halo). 

Our source is usually too faint for useful observations with the optical
 monitors of either {\sl Swift} or {\sl XMM-Newton}, it was out of their
 field of view in most exposures and
 in any case, source crowding and source confusion would also be problematic with these
 instruments.  However, we note that no luminous UV sources with magnitude
 approximately lower than 21 were observed in this field
 with either the {\sl Swift} optical monitor or with {\sl GALEX} \citep[][]{orio2010},
 ruling out very large variability of CXO J004318.8+412016.

The largely negative Q value and the the high optical-ultraviolet
 flux are the reasons for which we initially suggested
 that the true optical
 counterpart may be a very young and massive Be star, which
 is a possible classification based on the optical spectrum \citep[][]{orio2015}. 
 However, we have shown here that the SED is {\it not} consistent with a Be star,
 but only with a red giant.
 Symbiotic stars were given their name because they present the spectral
 blend of a very hot (UV-emitting) and a luminous red object (a 
 red giant or AGB, exceptionally perhaps a supergiant in case of
 a neutron star or black hole companion).
 The U luminosity in most cases is due to the symbiotic nebula, and possibly
 also to the accretion disk. However, the optical-UV luminosity (U, and F336W filters) of 
CXO J004318.8+412016 seems to be unusually high with respect to
 the red band luminosity. The 
 optical-UV flux cannot be attributed to the Raleigh-Taylor
 tail of the SSS, because the source temperature is so high, that
 90\% of the bolometric luminosity would be emitted in X-rays,
 with about 10\% of the remaining flux in the extreme UV \citep[see][]{orio2010}.

In the first (upper left panel) of Fig. 5 we show 
 that with a value E(B-V)=0.20, consistently with Fig. 4, the optical colors
  are well fit with the SED of a red giant
 of 1 M$_\odot$, T$_{\rm eff}$=5300 K, and a radius of 43 R$_\odot$ 
 (the same is true for the IR colors, as already demonstrated in Fig. 4).
 What causes the unusual excess in the ultraviolet bands?
 The possibility of an unresolved overlapping 
 object contributing to the ``weird'' optical-UV color 
 is very unlikely, given that it should be very luminous only
 in this band and not nearly as luminous in the nearby UV range. 
We also rule out a significant contribution of a symbiotic nebula,
 given that nebular lines are weak or absent.

 We also compared the average magnitudes and color indexes
 to a model of a disk in a binary, shown in the other panels of Fig. 5. 
We added an accretion disk to the SED of a secondary star, assuming both
 a red and  blue giant and varying the star's temperature as a fitting parameter
 (of course the presence of the disk implies a less luminous stellar component).
The accretion  disk was modeled according to \citet[][]{Patruno08}.
 With this composite fit, we first ruled out that a B or other luminous main sequence
 star can be consistent with the SED of our object if we add a disk, 
 because an accreting disk has a relatively flat spectral distribution.
We show here the red giant+disk fits in the region from the U to the I bands, the range of the
 spectrum included in the model by \citet[][]{Patruno08},
 and in which we have an unusual SED. 
 We performed several fits over a grid with
 different values of orbital periods, in each fit using
 the albedo and inclination as free parameters.
 The top right panel shows the fit adding to the red giant 
  an accretion disk around a WD of $1.3 M_\odot$,
accreting from a 1 M$_\odot$ donor with a radius of 43  R$_\odot$
 and T$_{\rm eff}$=4000 K in a 90 days orbit, at inclination
  80$^{\rm o}$ and albedo=0.8, and E(B-V)=0.20. In all fits, we 
 accounted also for irradiation of the secondary. 
In the bottom left panel, we found an approximate fit with a disk around a black hole of 5 M$_\odot$
from a donor star of 1 M$_\odot$, a radius of 30 R$_\odot$,
 temperature of 4000 K, inclination of 0$^{\rm o}$,
 albedo=0.95 (bottom left panel) and E(B-V)=0.20 
 in a 60 days period (bottom left).
 Finally, in the bottom right panel we show the fit with
 a disk around a 1.3 M$_\odot$ WD with a companion of 1 M$_\odot$,
 T$_{\rm eff}$=4000 K and a radius of 80 R$_\odot$
in a 230 days orbit, with inclination  0$^{\rm o}$,
albedo=0.5 and the higher absorption, E(B-V)=0.71.
  
 Assuming that the compact object is a
 WD, the disk hypothesis is quite consistent with the observed results,
 but there still is an excess towards the ultraviolet, which
 may be due to the Balmer jump in emission. Such a phenomenon is 
 observed in about 33\% of symbiotics \citep[][]{Munari2002, henden2008}
 although in the Galaxy only 4 WD-symbiotics
 show U-B$<$-1 (this is probably because most known Galactic symbiotics
 are affected by high reddening). 
 The larger absorption, E(B-V)=0.71 like we observed
 in our 2015 optical spectrum, is consistent with a low inclination,
 and with observing X-rays from a luminous central object (i.e., a WD).
We caution that the disk model is calculated only assuming that the secondary
 fills its Roche lobe, which is not the case in several
 observed WD symbiotics, even those that show evidence of an accretion disk. 
This is the reason the fit on the right side of Fig. 5 is obtained 
 with  short orbital periods of 60 and 90 days, not observed in symbiotics, 
 in which the average orbital periods are of the order of 2 years
 \citep[][]{Belczynski2000, miko2012}.
 Thus, these orbital periods should be regarded only as lower limits.
 
  Accretion disks appear to be common in WD-symbiotics \citep[see][and references
 therein]{nunez2016}, although
 \citet[][]{Kenyon83, miko2012}, among others, have shown that the orbital
 separations of WD-symbiotics are too large for Roche-lobe overflow,
 unless the secondary is deformed.
\citet[][]{miko2012} has pointed at the observed ellipsoidal
 variation of many WD-symbiotics
 as proof of disk formation in a modified Roche potential.
In addition, disks in WD-symbiotics may be formed  without
 Roche lobe overflow of the secondary,
if the red giant wind carries angular momentum
\citep[similarly to the model of][]{huarte2013}. Such a disk may be
 truncated and appear ``redder'' than the standard disk we assumed in Fig. 5.

  As shown in the bottom right panel of Fig. 5, 
 the rise towards the ultraviolet cannot be explained at all with
 a disk around a black hole, while it seems marginally consistent
 with a disk around a WD. 
  However, this is not a proof against the black hole hypothesis,
 because this part of the spectrum may contain a strong Balmer jump in emission.
 To summarize, the available photometric measurements do indicate quite clearly that 
 the binary hosts red giant, but we do not have sufficient data to really rule out
 a black hole central object. The X-ray characteristics are more typical
 of hydrogen burning on a WD, but without high resolution
 high energy spectra, which at present cannot be obtained yet at M31 distance,
 also the black hole cannot be ruled out. The solution
 may be given by measuring radial velocities
 of lines that may be emitted near the compact object. 

\begin{table*}
\centering
\caption{In this table, available in electronic format, we give the count
 rates we derived for the XMM-Newton observations, first the ones done 
 with EPIC-pn and the thin filter, then with EPIC-pn and the medium filter,
 both in the 0.2-1 keV and in the 0.2-1 keV ranges,
 and finally with EPIC-MOS and the medium filter in the 0.3-1 keV.
 The first column gives the modified Julian date, 
 the second column gives the net exposure times (once the solar
 flares or other bad intervals were removed), the third the total duration of
 the exposure, and
 columns 4 to 6 list the count rates and
 their errors (see text).} 
\label{tab:count rates}
\begin{tabular}{lcccccr} %
\hline 
Julian Date & Net exp. (s) & Exp. (s) & cts s$^{-1}$ (0.15-1 keV) &
error(0.15-1) & cts s$^{-1}$ (0.2-1 keV) & error(0.2-1) \\
\hline
\multicolumn{7}{c} {EPIC-pn, thin filter} \\
52281.28104167 & 55330 & 64317 & 1.039e-01 &  1.589e-03 & 5.767e-02 & 1.168e-03 \\
53896.10890046 &  3831 & 21913 & 5.507e-03 & 1.626e-03  & 6.826e-03 & 1.913e-03 \\
54100.68317130 &  12224 & 15918 & 2.284e-02 & 2.388e-03 & 5.046e-02 &  3.192e-03 \\
\hline
\end{tabular}
\end{table*}
\begin{table*}
        \centering
        \caption{Spectral parameters for the atmospheric and the
 blackbody model, with the 90\% confidence
 level errors, and the $\chi^2$ per degrees of freedom
 statistical parameter, during the exposure
 done with XMM-Newton on 2000 June 25, and with Swift on 
 2015 July 26. The net exposure time used
 to extract the spectrum was 22250 s for the pn,
 22450 s for the two MOS, 18250 s for the Swift XRT. L$_{\rm X}$, absorbed
 and unabsorbed, is derived from the flux in the 0.2-1 keV range
 obtained in the fit, for a distance of 783 kpc or distance
 modulus 24.45 \citep[][]{dalcanton2012}.
 The error on the flux or luminosity is 
 calculated assuming fixed N(H) and T$_{\rm eff}$, and the
  luminosity is expressed  
 in units of 10$^{37}$ erg s$^{-1}$. The 90\% confidence level contours
 for the bolometric luminosity
 in the blackbody fit, and for the  flux in the atmospheric
 fit, are unbound for the 2015 data. }
\label{tab:X-ray spectra}
\begin{tabular}{lcccr} %  columns, alignment for each
                \hline
     & Atm. (2000) & Bbody (2000)  & Atm. (2015) & Bbody (2015) \\
   \hline
N(H) (10$^{21}$ cm$^{-2}$) & 1.37$\pm$0.21 & 2.47$^{+0.59}_{-0.24}$ & 1.90$^{+0.80}_{-0.17}$ & 2.80$^{+3.70}_{-1.90}$\\
 & & \\
T$_{\rm eff}$ (eV)     &  86$\pm$4 &  67$^{+3}_{-6}$  &   67$^{+60}_{-31}$ & 46$^{+30}_{-40}$  \\
 & & \\
T$_{\rm eff}$ (K)      & 10$^6 \pm 5 \times 10^4$  & 7.78$^{+0.4}_{-0.8} \times 10^5$ & 7.74$^{+7.74}_{-2.41} \times 10^5$ &
 5.34$^{+3.47}_{-4.00} \times 10^5$ \\
 & & & & \\
L(bol) &  ---      & 61.3$^{+80.3}_{-29.7}$ & --- & 283.9  \\
 & & & & \\
L$_{\rm X}$(abs.) &  1.26$_{-0.28}^{+0.38}$ & --- & 0.43 & --- \\ 
 & & \\
L$_{\rm X}$(unabs.) & 7.09$_{-1.56}^{+1.84}$ & --- & 5.55 & --- \\
 & & \\
$\chi^2$ & 1.2 & 1.0 & $\leq$1.0 & $\leq$1.0 \\
\hline
        \end{tabular}
\end{table*}
\section{The X-ray data in the last 15 years}
The field of CXO J004318.8+412016 was observed numerous times with {\sl Chandra}, {\sl XMM-Newton}
 and {\sl Swift} in the last 15 years, however, many {\sl Chandra}-ACIS
 observations and the vast majority of the {\sl Swift} ones are too
 shallow for detection of the source, even in its high state. The 
 upper limits through non-detections in X-ray observations
 are quite higher than actual measurements at minimum,
 so we do not include them in Table 3. It turns out
 that XMM-Newton can observe M31 only from the end of December to mid-February,
 and then again for a short period in the Summer (July-August).
 As we reminded above, in \citet[][]{orio2006}, aspects of the 
 the X-ray variability of the source were
 discussed used the available data at that time; 
 the column density did not seem to increase with decreasing flux,
 but probably the opposite was true, suggesting that when
 the luminosity increases above a certain level, a depleting wind prevents
 the source from exceeding the Eddington luminosity. 

 The long term X-ray light curve of  CXO J004318.8+412016 is shown in Fig.6,
 with data obtained from the EPIC cameras of {\sl XMM-Newton}, EPIC-pn
 with either the thin or the medium filter and the MOS 1 and MOS 2 with
 the medium filter, and with the Chandra HRC-I, which is very sensitive
 in the very soft range. Finally, three significant {\sl Swift} XRT
 detections were included for 2015-2016 (the previous {\sl Swift} exposures
 were too short for a detection, or our source was at the 
 very edge of the field).  All the count rates have been converted to the
 EPIC-pn/thin-filter count rate in the
 0.2-1 keV range (there are no significant
  counts above 1 keV), using the WebPIMMS on-line tool in HEASARC
 (FTOOL PIMMS v4.8b) and assuming a 
 blackbody with a temperature of 70 eV and a column density
 N(H)=2 $\times$ 10$^{21}$ cm$^{-2}$, taken
 as average characteristics.
 As an indication, a count rate of
 0.1 cts s$^{-1}$ measured with EPIC-pn and the thin filter
 in the 0.2-10.0 keV range
 translates in an absorbed flux of 9.6 $\times$ 10$^{-14}$ erg s$^{-1}$
 in the same range, with the above model.
 For this conversion
 we thus assumed the simplistic approximation that the spectrum does not vary,
 in order to give a term of comparison among the different 
 instruments. In several observation the spectrum is measured
 at low S/N and there is a large uncertainty in the best fit parameters,
 so we looked for an approximate comparison and not an exact one.
 In any case, the flux fluctuations  are much larger than uncertainties
 in the conversion between different observations and instruments,
 even with possible spectral variation. 

 Like in the observations done until 2005 \citep[][]{orio2006},
 we find that most observations indicate quite higher 
 column density than between us and M31. In Fig. 7. we show fits with
 an atmospheric model to an observation done on 2000 June 25,
 one of the dates of largest X-ray flux of our source, and
 to an observation done on 2015 July 26, 15 years later and close
 to the Gemini observation.
 Table 4, available on line, reports all
 the count rates for the positive detections: the {\sl XMM-Newton} ones 
 were obtained with the XMM-SAS  version 15.0.0 and its tool XSELECT,
 the {\sl Swift} X-ray telescope (XRT) ones with the online tool
 of the UK Swift Data Center, 
 while the Chandra HRC-I count rates were measured by \citet[][]{hofmann2013}.
 There is no clear evidence that the spectrum softens
 in the lower states, although most of the low-luminosity observations
 are not of sufficient good quality to obtain  statistically very meaningful spectral fits.  

 In Fig. 7 we show an example for a high X-ray flux period:
 fitting the observed spectrum in an {\sl XMM-Newton} exposure
 of June 25 2000 (observation 0112570401) with
 a blackbody indicates  super-Eddington luminosity for
 a stellar mass of less than 4.9 M$_\odot$, at the 90\% confidence level.
 On the
 other hand, the fit to the spectrum with a WD atmospheric model \citep[][]{rauch2010}
 indicates a luminosity of 7.1 $\times$ 10$^{37}$ erg  s$^{-1}$
 in the X-ray range, corresponding to more than 90\% the bolometric
 luminosity. Therefore, if the X-ray source is a nuclear burning WD, 
 it is not emitting super-Eddington luminosity. We remind that
 a  blackbody fit overestimates 
 the luminosity and underestimates the temperature of a hydrogen
 burning WD atmosphere, \citep[see][]{rauch2010, ness2011}. 
The T$_{\rm eff}$ derived from the atmospheric fit is 86 eV,
consistent with a WD mass between 1.2 and 1.3 M$_\odot$ according
 to \citet[][]{wolf2013}.
 
 We plotted the fit with the two different models in Fig. 7, and
 it is clear that it is very difficult to establish which one is more
 appropriate, even if
 the blackbody fit yields a lower value of $\chi^2$,
 1 versus 1.2. The atmospheric fit is much more complex and requires 
 fine tuning good quality data, so we do not consider this as proof that 
 a blackbody is a much better fit, as expected if an accretion disk rather than
 an atmosphere is the origin of the X-rays. We also note that the fit to
 the 2015 data results in higher values of column density than
 in the 2000 observation, but given the uncertainties, this probably does not
 indicate a trend towards higher absorption over the years. In fact,
 the luminosity in the optical-ultraviolet band is higher
 in the PHAT measurement obtained in 2013 than in the LGS images of
 2001-2002, which argues against a long-term absorption increase. 

 Although with the data at hand we cannot rule out
 that the X-ray luminosity variations are aperiodic, there is a possibility
 that we are observing some kind of periodic obscuration due to
 a wind that is optically thick to soft X-rays and is observed
 only at a given orbital phase, like in the symbiotic star and
 supersoft X-ray source SMC 3 in the SMC \citep[][]{orio2007,
sturm2011, kato2013}.
 We detected no clear variability during the  single exposures, neither between 
exposures repeated after few hours or 
 a day (see inset in Fig.5, showing exposures repeated
 for 4 consecutive days).  We also rule out variability with
 an amplitude of more than $\simeq$15\% on time
 scales of hours. Power spectra of the 
 soft X-ray EPIC-pn light curve in the two longest observations (close to
 7 hours) done while the source was 
 in a ``high'' state revealed no significant peaks. There
 is also no evidence that the X-ray luminosity variation on
 time scales of weeks may be
 due to an eclipse, because as we see in the inset of the
 second panel the low state can last for several days
 with a rather flat light curve, unlike for SMC3 
\citep[][]{kahabka2004, sturm2011}
 where there is a sharp drop and rise of flux.
 We cannot rule out also periodic modulations 
 shorter than about 6 months and, probably, longer than about 3 weeks,
 because this is the time it takes for a dimming and re-brightening
 of the source in 
 several XMM-Newton observations, with repeated
 fluctuations from the low to the high state always repeated, even after
 several years.
 A {\sl Swift} XRT exposure done in 2015 indicates high luminosity on 
 09/14/2015, very close in time to our Gemini exposures.  

\begin{figure*}
 \includegraphics[width=120mm]{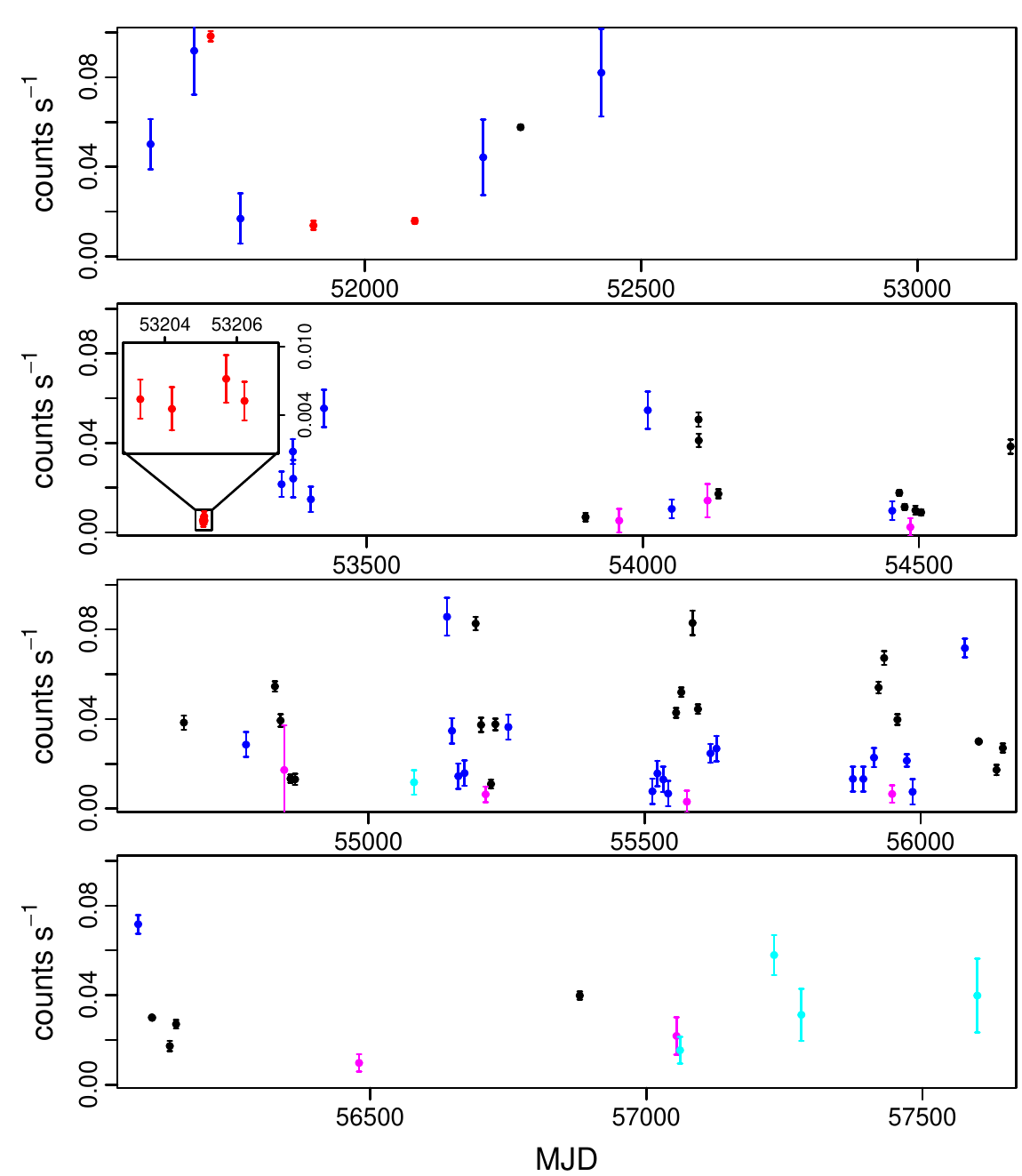}
 \caption{X-ray light curve obtained with {\sl XMM-Newton} EPIC-pn and thin
 filter (black), EPIC-pn and medium filter (red), EPIC-MOS and the
 medium filter (purple), {\sl Swift} XRT (light blue),
 and the Chandra HRC-I (blue). Because of lack of significant signal above
 1 keV, in order to reduce the noise
 the original count rates were extracted in the 0.2-1 keV range with EPIC-pn, 
 in the 0.3-1 keV range for EPIC-MOS and {\sl Swift} XRT.
 We used the count rates of the Chandra HRC-I from \citet[][]{hofmann2013}
 and converted all count rates in 
 equivalent EPIC-pn thin filter count rate in the 0.2-1 keV range,
 assuming the model described in the text. 
 The inset the second panel shows a zoom of observations obtained in 4 consecutive
 days, to show that the variability time scale is longer than
 few days.} 
\end{figure*} 
\begin{figure*}
 \includegraphics[width=85mm]{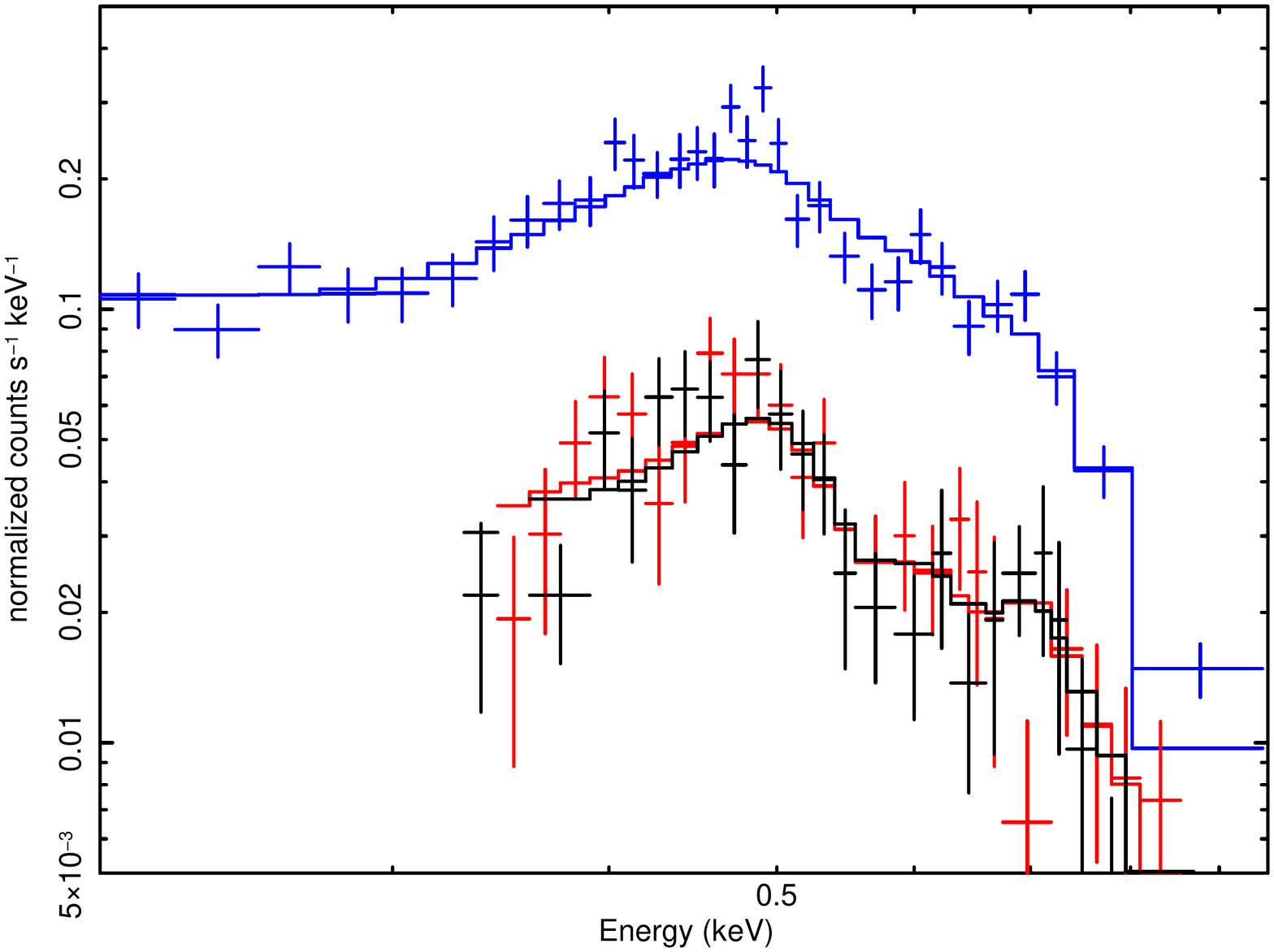}
\includegraphics[width=85mm]{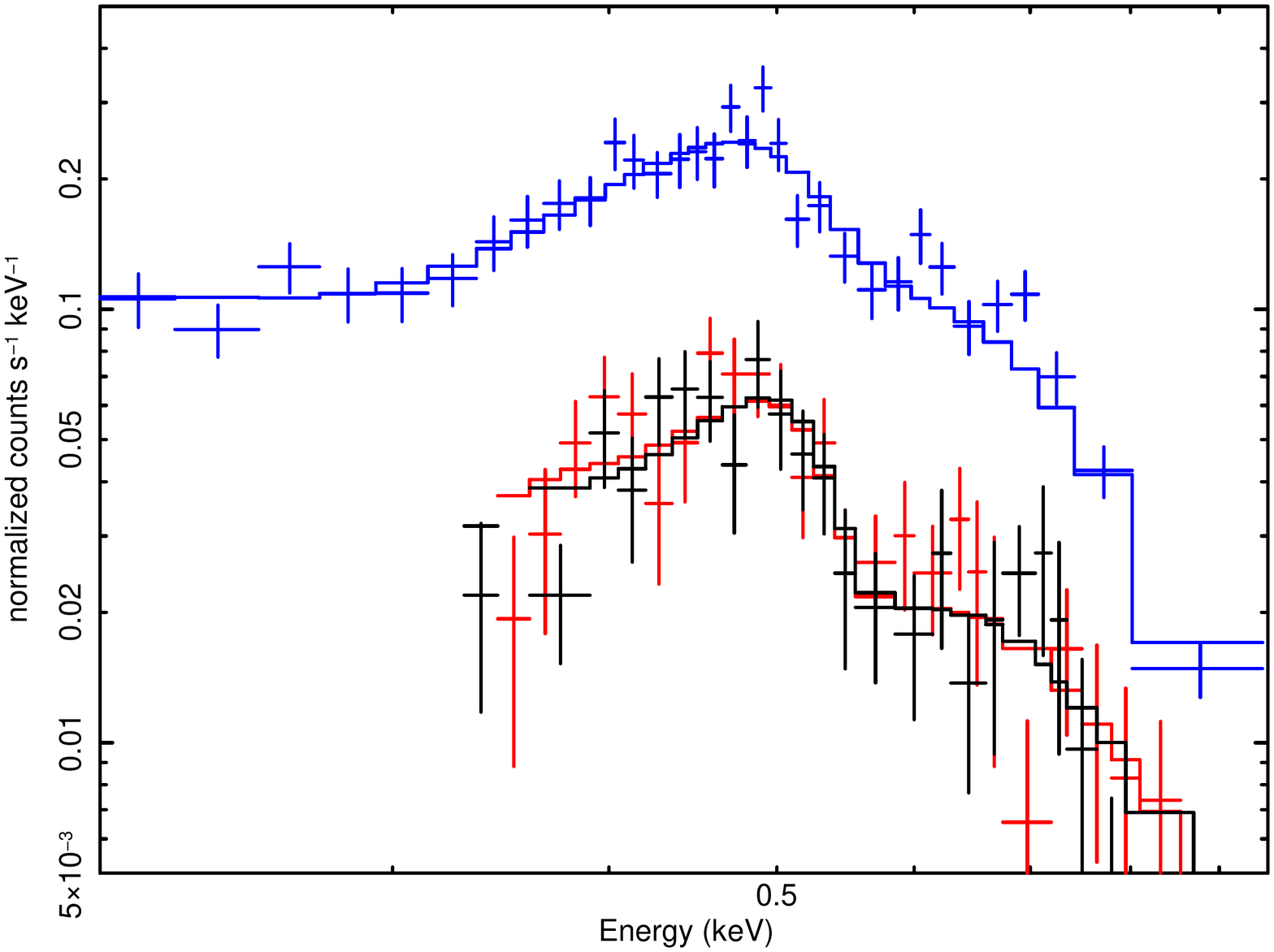}
 \includegraphics[width=85mm]{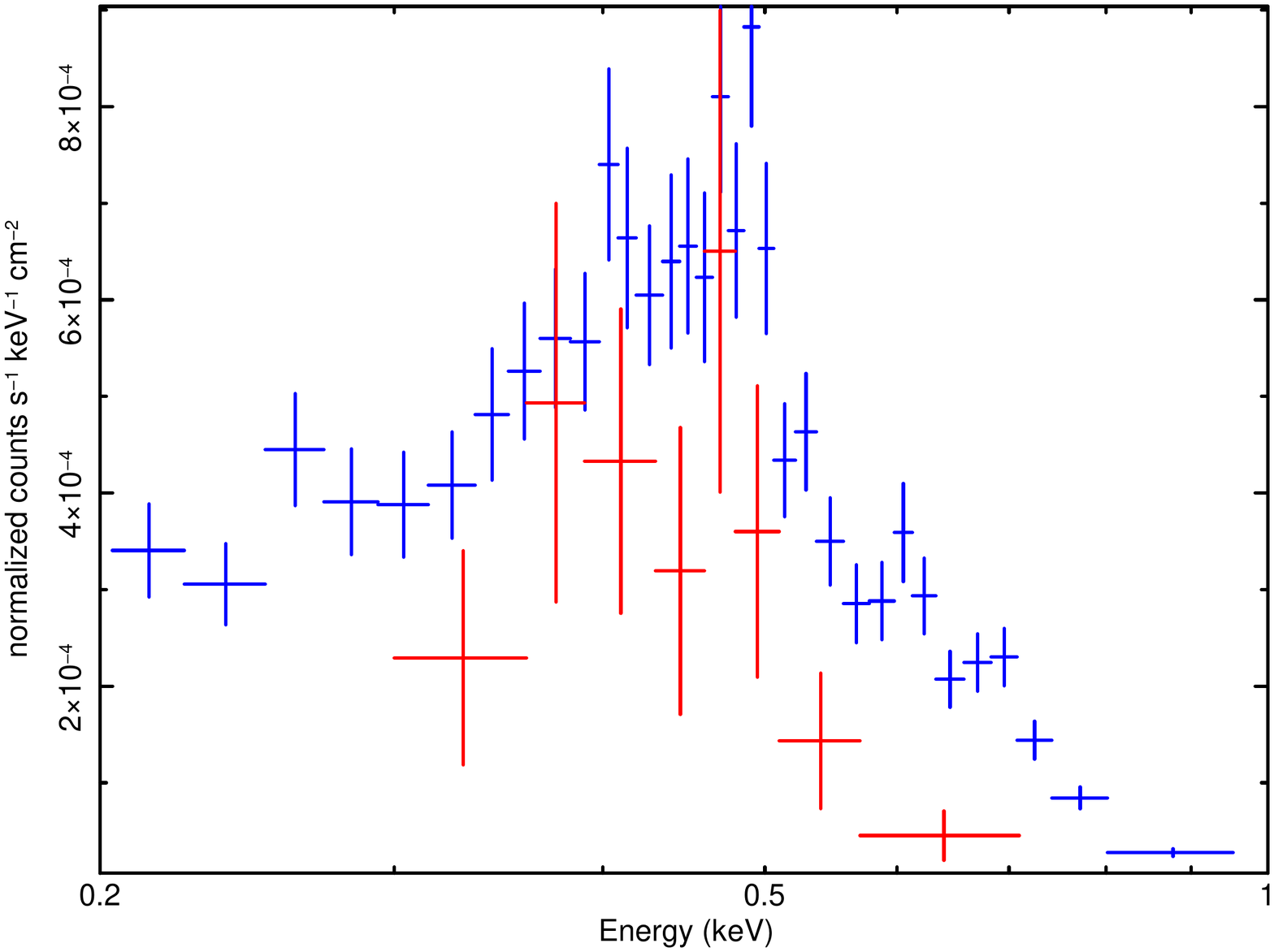}
 \includegraphics[width=85mm]{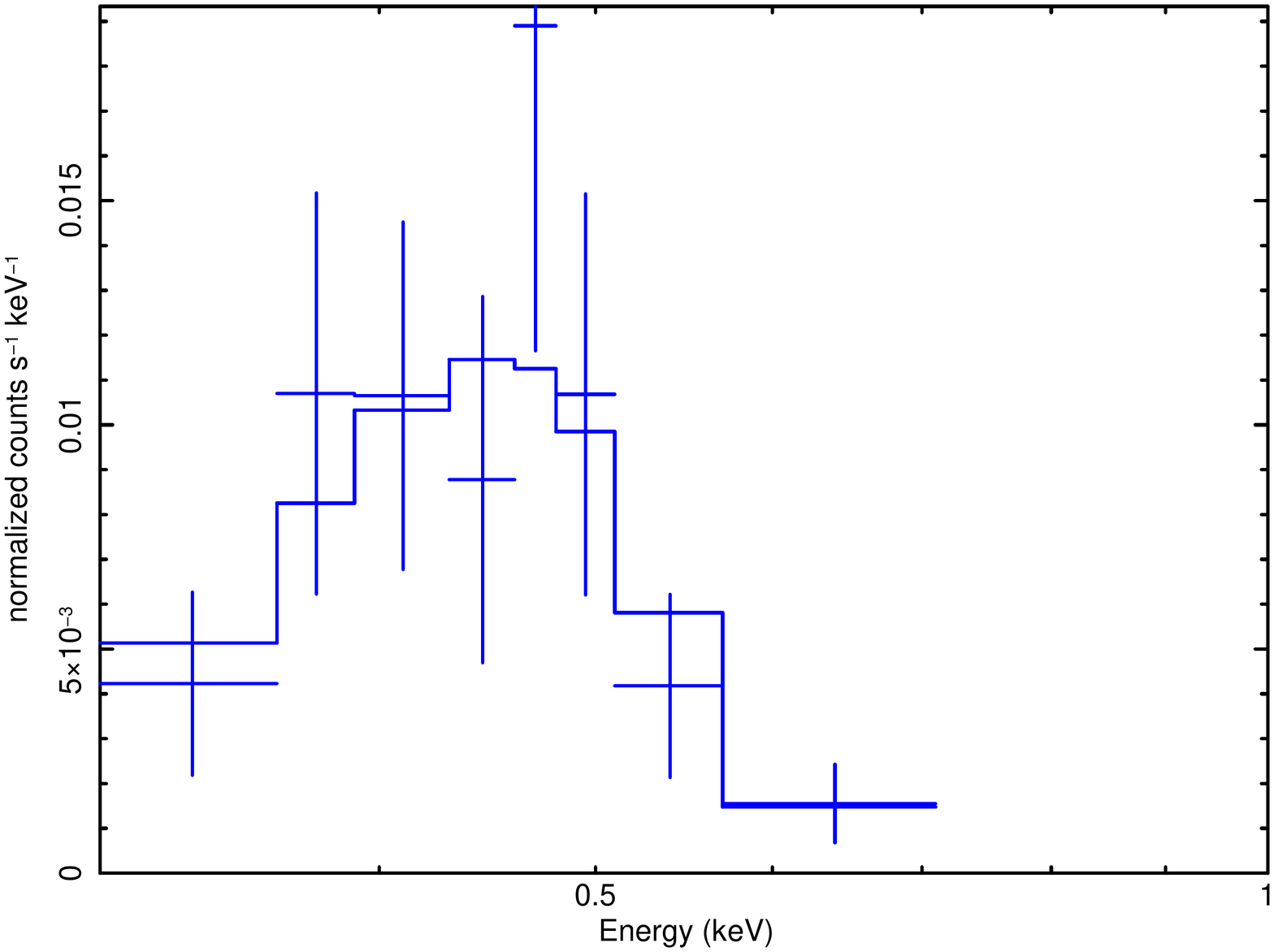}
 \caption{The upper panels show the count rate spectrum of the X-ray source observed with XMM-Newton EPIC-pn (blue) and 
with EPIC MOS-1 (black) and MOS-2 (red) on 2000 June 25 (this is the
 observations obtained with the highest S/N), and a fit
 with a WD atmospheric model on the left, with a blackbody on
 the right(see Table 4). In the lower panel 
 on the left, we compare the EPIC-pn spectrum of 2000 June 25 with
 the Swift-XRT spectrum of the source observed on 2015 July 26, both
 convoluted with the effective area and thus plotted in units of counts/s/cm$^{-2}$ (left). On
 the bottom right panel, the Swift-XRT spectrum of the same date fitted with a blackbody, 
 as in Table 4.}
\end{figure*}
\section{Some crucial aspects}

The following puzzling characteristics of this source,
 deserve further observations and related modeling:
 
1) Given the high effective temperature of the X-ray source,
 coronal lines of [Fe X], especially 
 the one at 6374 \AA, which is very strong in the spectrum of SMC 3,
 should have been detected in the spectrum of CXO J004318.8+412016.  
 If the X-rays are due to hydrogen burning, either hydrogen burning had temporarily been
 shut off when we
 observed it with Gemini, or CXO J004318.8+412016 is surrounded by much denser material
 than the symbiotics we know host a very hot WD. 
The electron density to avoid the [Fe X] forbidden transition
 is above a critical value 
 of about 5 $\times 10^9$ cm$^{-3}$ \citep[e.g.][]{nagao2002, kato2013}.
 We also note that
 the ratios of the intensity He I line at 6678 \AA \ to those of  the He I line
 at at 7065 \AA \ or the He I triplet
 at 5876 \AA  \ indicate higher electron density than $10^7$ cm$^{-3}$.
 Although this electron density appears to be higher than often observed
 in WD-symbiotics, this is not 
 evidence against the WD presence. In fact 
 all symbiotics, especially those with non-Mira donors, have 
 at least some regions of high density. This is indicated, for
 instance, by   
their optical He I singlet to triplet ratios, [OIII] to H I line ratios, 
and UV intercombination line ratios. 
This fact is even used for distinguishing between symbiotic stars and
planetary nebulae, like in
 the  [OIII] diagnostic diagram used by \citet[][]{Gutierrez1999, miko2014} 
in the He I diagram of \citet[][]{Proga1994, miko2014}.
 
2) Another characteristic of the optical spectrum is the large Balmer decrement,
 that we attribute to an optically thick medium.
 If it is due to intrinsic reddening in the binary, it indicates E(B-V)$\simeq$0.71,
 more than three times higher than what we infer from the
 dust maps of \citet[][]{dalcanton2015}. {\sl Swift} XRT exposures
 yielded detections were done within $\simeq$3 weeks before and
 after Gemini observations, but although the source may 
 have varied within this time,
 the data are consistent with N(H)$\simeq 4 \times$ 10$^{21}$ cm$^{-2}$
 and with the measured Balmer decrement. We stress that there 
 is no simultaneous X-ray observation, but we know 
 that the fit to the X-ray spectrum indicates that at least in many
 of the X-ray observations
 the value of the column density was lower. 
 Will we observe a varying Balmer decrement if we take optical spectra
 at other epochs? It would be an interesting prove of variable 
 absorption, due to a wind or other mass ejection phenomenon.

 3) The SED of this object is dominated by the presence of a red
 giant as the secondary. It is not an M giant, but of
 spectral type K or even earlier. 

4) Although we favor an explanation in terms of  a hydrogen burning WD,
 a black hole primary cannot be ruled out yet with the existing data. 

\section{Conclusions}
 The prominent emission lines of the luminous object we observed
 in the narrow spatial error circle
 of CXO J004318.8+412016 are typical of the 
 optical counterparts of X-ray binaries in general, and of supersoft X-ray
 sources more specifically. 
 According to the kinematics, it belongs to the M31 population.
 The continuum energy distribution from the IR to the UV is typical
 of an evolved star on the way to becoming a red giant, and 
 the additional spectral energy distribution of an accretion disk 
 only partially explains the rise of the flux towards the ultraviolet.
 This rise may be due to  the Balmer jump in emission, which must
 be significant in a source with  prominent Balmer lines. 

 The red giant SED that definitely suggests a symbiotic classification,
 but the optical spectrum presents several differences from 
 those previously observed in a few other known hydrogen burning WD-symbiotics. 
 Shell burning on a massive WD remains the most likely origin 
 of the high supersoft luminosity in
 this source but we cannot rule
 out a black hole binary. The high optical depth and  the likely possibility
 that the material from which the optical emission lines arises has
electron density n(e)$\geq 5 \times 10^9$ cm$^{-3}$,
indicate significant intrinsic  absorption. In
 the last 25 years, the absorption has been
 high and variable, although it
 was mostly not sufficiently high
 to absorb the supersoft X-rays of this very luminous source.
It would be important to determine the precise time scale of the X-ray variability
 and its possible periodicity. The X-ray light curve resembles
 the observed fluctuations known to occur in the X-ray luminosity
 of CAL 83, an SSS known as a WD burning 
 binary, with a main sequence or slightly evolved secondary
 of higher mass than the WD \citep[see][]{lanz2005}. Measuring a  period
 of the X-ray variability 
 would also allow us to understand whether the X-ray source
 had shut off at the time the optical spectrum was
 taken: could this be the reason of the missing, or weak, emission lines due to 
 high ionization or high excitation transitions?
 {\sl Swift} XRT exposures about 3 weeks before and after the date on which almost  
 all stacked spectra were obtained with Gemini, allowed
 us to measure a still active SSS. So, the time scale
 for a temporary shut-off of the burning would be of the order 
 of only few weeks. 

 The X-ray luminosity of CXO J004318.8+412016 oscillates between a few
 times 10$^{35}$ erg s$^{-1}$ and  
 a few times 10$^{37}$ erg s$^{-1}$, in the range of hydrogen burning
 WDs, as observed in post-outburst novae. 
 We confirm that with the data at hand, it seems unlikely that this
 variation is related to changes
 in the absorption column. On the contrary,
 the variation  of absorption column may be anti-correlated
 with the luminosity \citep[][]{orio2006}. 
 A tantalizing idea is that the supersoft X-ray luminosity variations 
 indicate instead semi-degenerate thermonuclear flashes, repeated on time
 scales of less than 3 months. When even a modest amount
 of material is ejected, the optical and X-ray luminosity
 increase, even if there is larger intrinsic absorption. In recent years, a nova has been
 observed to outburst in M31 with a recurrence period of less than a year
\citep[][]{henze2015, darnley2016}. With even shorter recurrence times,
 the flash is predicted to occur in only mildly degenerate
 conditions, with a luminosity increase of small
 amplitude compared with known classical
 novae, and almost without mass loss \citep[see][]{fuji1982, wolf2013}.

Although a very interesting group of SSS are found in young, massive binaries
 in the Magellanic Clouds \citep[][and references therein]{orio2013},
the majority of the persistent SSS we know are in symbiotic binaries.
 If the X-rays are due to hydrogen burning on a WD,
  because of the high $\dot m$ required
 for  persistent burning, we can speculate that during the red giant phase of the
 secondary, some mechanism accelerates mass transfer.
 Also recurrent nova outbursts, which
require high $\dot m$, but less than an order of magnitude than needed
 for steady burning (about 10$^{-8}$ M$_\odot$ year$^{-1}$)
 seem to be common in WD-symbiotics. 4 WD symbiotics that exploded as recurrent novae are
 know in the Galaxy, out of 16 known WD-symbiotics observed to undergo
  thermonuclear runaways \citep{miko2012}.
 Moreover,  several observations indicate
 that surface hydrogen burning almost always occurs in the WDs of symbiotic systems
 \citep[][and references therein]{miko2012}.
 \citet[][]{luna2013} found a fraction of symbiotics
 with no detectable fast UV variability, suggesting that their luminosity
 is powered by nuclear burning on low mass WDs.

These facts have made  WD-symbiotic appear very interesting as SNe Ia
 progenitors as single degenerate binaries. 
 However, proving or ruling out
 that they are a significant channel to SNe Ia
while they are still single degenerate systems (we note that double degenerates
 also need to have a symbiotic evolutionary phase), requires much
 better statistics than we currently have.  The  SSS would
 only rarely be detectable, because of the large intrinsic
 absorption of the symbiotic's nebula and wind, so the observed
 SSS are only
 the ones effected by low absorption, that is, the tip of the iceberg.
 The sample of SSS WD-symbiotics in nearby
 galaxies includes one likely symbiotic nova, RX J0550.0-7151
 \citep[][]{schmidtke1995, charles1996} and
 three persistent sources,  SMC 3, Lin 358 and Draco C-1, the last two
 at low T$_{\rm eff} \leq$200,000 K.
 We now have the means to study symbiotics at large distances
 in the Local Group.
 \citet[][]{miko2014} have identified 35 symbiotics in Andromeda,
 \citet[][]{goncalves2006} discovered one in IC 10 (2008),
 \citet[][]{goncalves2015}
 presented a symbiotic and
 two additional candidates in NGC 205, \citet[][]{kniazev2009} discovered
 one in NGC 6822, 
and 12 are known in M33 \citep[][]{miko2017}. These objects
 were found through H$\alpha$ imaging. Due to the
 detection limits, we estimate that they probably represent only the
 20\% most optically luminous symbiotics at $\simeq$800 kpc distance.
 Several authors noticed that they mostly have AGB companions.

 We suggest that coordinated X-ray and optical observations of 
 CXO J004318.8+412016 should be done 
 in the near future. First of all, it would be important to follow
 the variations of optical depth and/or intrinsic absorption (from the optical
 spectrum) and column
 density N(H), assessing whether they are correlated.
Other than the temporary shut-off of the burning, 
 there is a possibility that the emission lines corresponding
 to the transitions with the highest ionization potential were
 not observed in the optical spectrum because
 of a peculiar geometry and distribution of the absorbing gas
 in the system.  Ultimately, it would
 be extremely interesting for the evolutionary
 models to understand whether the brightening and
 dimming of the source is due to a nova-like phenomenon:
 this would be a very fast recurrent nova of very small amplitude, possibly
 an only mildly degenerate thermonuclear runaway, without mass outflow.

 Another important set of observations should aim at the measurement of the
 radial velocity displacement of the
 emission lines that may be produced near the compact
 object, so with some 
 assumption we should be able to estimate the mass of the compact object,
and assess whether  it is indeed a massive WD. We are still
 unable to rule out another interesting possibility for the 
 nature of the object, that of a black hole binary.
 Spectroscopic monitoring over a few years may reveal 
 radial velocity displacements of the emission lines,
 which will be crucial to assess the nature of this 
``extreme'' and intriguing X-ray source.
 
\section*{Acknowledgements}
G. Juan Luna is a member of the CIC-CONICET and his work was supported 
by grant Pip-Conicet/2011 \#D4598, ANPCYT-PICT 0478/14.
Ralf Kotulla  gratefully acknowledges financial support from
 the National Science Foundation under Grants
No. 1412449 and No. 1664342.
Joanna Mikolajewska acknowledges support of the Polish National Science
 Centre grant DEC-2013/10/M/ST9/00086,
and Domitilla de Martino of ASI-INAF grant I/037/12/0.

%%%%%%%%%%%%%%%%%%%%%%%%%%%%%%%%%%%%%%%%%%%%%%%%%%



\bibliographystyle{mnras}
\bibliography{biblio} % if your bibtex file is called example.bib
%%%%%%%%%%%%%%%%%%%%%%%%%%%%%%%%%%%%%%%%%%%%%%%%%%

%%%%%%%%%%%%%%%%% APPENDICES %%%%%%%%%%%%%%%%%%%%%

\appendix

%%%%%%%%%%%%%%%%%%%%%%%%%%%%%%%%%%%%%%%%%%%%%%%%%%

% Don't change these lines
\bsp	% typesetting comment
\label{lastpage}
\end{document}